\documentclass[10pt,twoside,english]{iopart}
\usepackage[T1]{fontenc}
\usepackage[latin9]{inputenc}
\usepackage{xcolor}
\usepackage{babel}

\usepackage{graphicx}
\usepackage{amssymb}
\usepackage{esint}
\usepackage[numbers]{natbib}
\PassOptionsToPackage{normalem}{ulem}
\usepackage{ulem}
\usepackage[unicode=true, pdfusetitle,
 bookmarks=true,bookmarksnumbered=false,bookmarksopen=false,
 breaklinks=false,pdfborder={0 0 1},backref=false,colorlinks=false]
 {hyperref}

\makeatletter

\providecolor{lyxadded}{rgb}{0,0,1}
\providecolor{lyxdeleted}{rgb}{1,0,0}

\usepackage{iopams}
\usepackage{setstack}

\usepackage{braket}

\makeatother

\begin{document}

\title{Creation and dynamics of remote spin-entangled pairs in the expansion
of strongly correlated fermions in an optical lattice }

\author{Stefan Ke{\ss}ler$^{1}$, Ian P. McCulloch$^{2}$, and Florian Marquardt$^{1,3}$}

\address{$^{1}$Institute for Theoretical Physics, Universit\"{a}t Erlangen-N\"{u}rnberg,
Staudtstr. 7, 91058 Erlangen, Germany\\$^{2}$Centre for Engineered
Quantum Systems, School of Mathematics and Physics, The University
of Queensland, St Lucia, QLD 4072, Australia \\ $^{3}$Max Planck
Institute for the Science of Light, G\"{u}nther-Scharowsky-Stra{\ss}e
1/Bau 24, 91058 Erlangen, Germany}
\begin{abstract}
We consider the nonequilibrium dynamics of an interacting spin-$\frac{1}{2}$
fermion gas in a one-dimensional optical lattice after switching off
the confining potential. In particular, we study the creation and
the time evolution of spatially separated, spin-entangled fermionic
pairs. The time-dependent density-matrix renormalization group is
used to simulate the time evolution and evaluate the two-site spin
correlation functions, from which the concurrence is calculated. We
find that the typical distance between entangled fermions depends
crucially on the onsite interaction strength, and that a time-dependent
modulation of the tunneling amplitude can enhance the production of
spin-entanglement. Moreover, we discuss the prospects of experimentally
observing these phenomena using spin-dependent single-site detection. 
\end{abstract}


\maketitle
\tableofcontents{}

\section{Introduction}

\subsection{Motivation}

The tremendous experimental progress with ultra cold atoms in optical
lattices makes them a unique playground for studying nonequilibrium
dynamics of many-body systems. The remarkable features of these systems
are the precise and dynamical control of the interparticle interaction
and external trapping potentials, as well as long coherence times
\citep{Bloch2008}. Different nonequilibrium initial states can be
generated by a quantum quench (for a recent review see \citep{Polkovnikov2011}):
a parameter in the Hamiltonian is suddenly changed such that the system,
initially in the ground state of the Hamiltonian, is afterwards in
an excited state of the new Hamiltonian. Quenches have been experimentally
realized, for instance, in the interaction strength \citep{Greiner2002b,Widera2008,Chen2011b}
and in the additional trapping potential by either switching it off
\citep{Schneider2012} or displacing its center \citep{Ott2004,Pezze2004,Strohmaier2007}.
This led to observations such as the collapse and revival of the coherence
in a Bose-Einstein condensate after a quench from the superfluid to
the Mott insulating regime \citep{Greiner2002b}.

New detection schemes \citep{Wuertz2009,Bakr2009,Sherson2010} provide
the possibility of observing the many-body state at the single-site
and single-atom level and make these systems even more suitable for
studying the dynamics of (spatial) correlations in nonequilibrium
situations. These methods have already been used to monitor the propagation
of quasi-particle pairs \citep{Cheneau2012,Barmettler2012} and a
single spin impurity \citep{Fukuhara2013} in a bosonic gas. They
have also inspired theoretical work on many-body dynamics subject
to observations, treating issues such as entanglement growth in a
bosonic system \citep{Daley2012}, destructive single-site measurements
\citep{Barmettler2011}, and Quantum Zeno effect physics for spin
\citep{Gammelmark2010} and expansion \citep{Kessler2012} dynamics.

In the present article, we study a different aspect of these systems:
the creation and dynamics of spin-entanglement during the expansion
of a strongly interacting, spin-balanced fermionic gas in an optical
lattice. We find that the expansion out of an initial cluster of fermions
can automatically generate long-range spin-entanglement. 

Expansion dynamics of interacting fermions in an optical lattice has
been realized recently in a three-dimensional optical lattice \citep{Schneider2012}.
In that experiment, the authors observed a bimodal expansion with
a ballistic and diffusive part and were able to explain its anomalous
behaviour using a Boltzmann-based approach. In addition, expansion
dynamics of this kind has been studied numerically in one dimension,
going beyond a kinetic description. These studies revealed the dependence
of the expansion velocity \citep{Kajala2011,Langer2012}, the momentum
distribution function, and the spin and density structure factors
\citep{Heidrich-Meisner2008} on the onsite interactions and the initial
filling. Furthermore, the effects of the different quench scenarios
\citep{Karlsson2011} and of the gravitational field \citep{Mandt2011}
on the time-evolution of the density distribution have been discussed.
The sudden expansion of a spin-imbalanced fermionic gas has been recently
considered for observing Fulde-Ferrell-Larkin-Ovchinnikov correlations
\citep{Kajala2011a,Lu2012,Bolech2012}.

In general, the dynamics is crucially affected by the difference in
behaviour between a single fermion and that of a doublon (i.e., a
pair of fermions at the same lattice site). For large interaction
strengths doublons are very stable against decay into fermions (analogously
to the repulsively bound boson pairs \citep{Winkler2006}) and move
slowly. These properties can lead to the condensation of doublons
\citep{Rosch2008} and a decrease of the entropy \citep{Heidrich-Meisner2009c}
in the center of the cloud during the expansion. They play also a
role in the decay dynamics of doublon-holon pairs in a Mott insulator
\citep{Al-Hassanieh2008,DiasDaSilva2010,DiasDaSilva2012}. In the
present work, we will show how spin-entanglement is generated by the
decay of a doublon into single fermions.

While we focus here on the build-up of correlations in a many-body
state (which is mainly driven by the creation of correlated fermions
out of doublons), the efficient production of entangled atom pairs
is by itself an important topic, especially in the context of atom
interferometry \citep{Cronin2009}. Using such nonclassical atom pair
sources would allow matter-wave optics beyond the standard quantum
limit. Recent experiments succeeded in generating large ensembles
of pair-correlated atoms from a trapped Bose-Einstein condensate,
using either spin-changing collisions \citep{Luecke2011,Gross2011}
or collisional de-excitation \citep{Buecker2011}. In the context
of such experiments, detection methods able to image single freely
propagating atoms have also recently become available \citep{Buecker2009}.

The article is organized as follows: First, we will describe in detail
the expansion protocol, starting from a finitely extended band insulating
state, i.e., a cluster of spin-singlet doublons. Moreover, we give
details about the numerical time-dependent density matrix renormalization
group (tDMRG) simulation. Before studying spin-physics, we first discuss
correlations in the density of the expanding cloud (section~\ref{sec:Density-density-correlation}),
where the influence of interactions is already significant. We find
that large onsite interactions lead to positive correlations between
certain lattice sites. The dynamics of spin-entangled pairs, the main
focus of our work, is then presented in section~\ref{sec:Spin-entanglement}.
First of all, we relate the concurrence to the spin-spin correlation
functions. Then we discuss the propagation of spin-entangled pairs
in the cloud. Furthermore, we compare the cumulative {}``amount''
of spin-entanglement in different regions of the cloud and for various
interaction strengths. We find that spin-entanglement between remote
lattice sites is only found for large interaction strength, while
for small onsite-interactions there is entanglement preferentially
only between nearby sites. Moreover, the Pauli-blocked core of the
cluster favours both partners of a spin-entangled pair to be emitted
into the same direction when compared to the decay of a single doublon,
where they almost always are emitted into different directions. In
addition, we discuss the expansion under the action of a time-dependent
(modulated) tunneling amplitude. We find that the modulation can enhance
the production of spin-entanglement. Finally, we discuss a possible
experimental setup for observing this spin-entanglement dynamics in
the near future using spin-dependent single-site detection (section~\ref{sec:Experimental remarks}). 

\begin{figure}
\begin{raggedleft}
\includegraphics[scale=0.28]{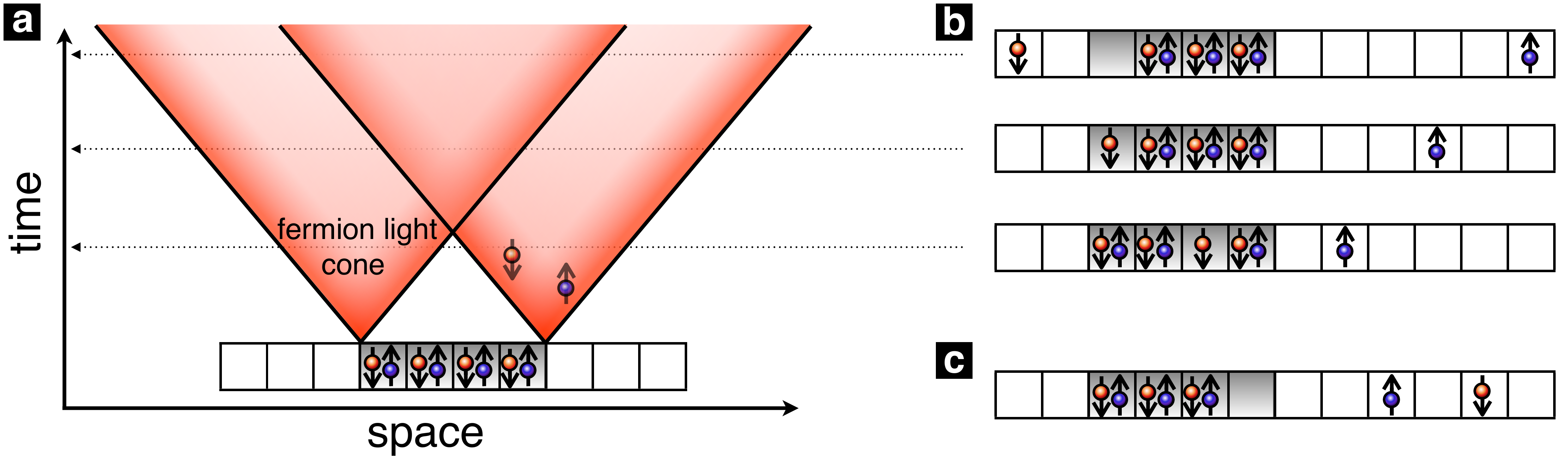}
\par\end{raggedleft}

\caption{Schematic of the expansion from a band insulating initial state into
an empty one-dimensional lattice. (a) Due to the discreteness of the
lattice the velocity of the fermions is bounded. Thus, the fermions
emitted from a doublon move within a light cone (shown for the leftmost
and rightmost doublon). (b,c) Examples of fermion configurations at
different evolution times, where only the rightmost doublon decayed
into single fermions and the other doublons remained at the original
position. (b) A spin up (down) excitation can propagate through the
band insulating cluster as a spin down (up) hole (time evolution from
bottom to top). (c) In this case, both fermions have been emitted
to the same direction. \label{fig: Schematic}}

\end{figure}

\subsection{Model}

In this paper, we consider an ultracold gas of fermionic atoms loaded
into a one-dimensional optical lattice. The atoms can be prepared
in two different hyperfine states, which we label $\uparrow$ and
$\downarrow$, and fermions of different {}``spin'' interact via
s-wave scattering. This fermionic mixture represents a realization
of the standard fermionic Hubbard Hamiltonian \citep{Schneider2008,Joerdens2008}

\begin{equation}
\hat{\mathcal{H}}=-J\sum_{i=1}^{L-1}\sum_{a=\uparrow,\downarrow}\left\{ \hat{c}_{i,a}^{\dagger}\hat{c}_{i+1,a}+\hat{c}_{i+1,a}^{\dagger}\hat{c}_{i,a}\right\} +U\sum_{i=1}^{L}\hat{n}_{i,\uparrow}\hat{n}_{i,\downarrow}.\label{eq:FH Hamiltonian}\end{equation}
The first term describes the tunneling of fermions between adjacent
lattices sites with amplitude $J$. The second term encodes the effective
onsite interaction energy $U$ of fermions with different hyperfine
states, which can be controlled using a Feshbach resonance. The particle
number operator is $\hat{n}_{i,a}=\hat{c}_{i,a}^{\dagger}\hat{c}_{i,a}$,
with the creation (annihilation) operator $\hat{c}_{i,a}^{\dagger}$($\hat{c}_{i,a}$)
satisfying the fermionic commutation relations. 

In the following, we focus on the expansion from a band-insulating
state, i.e., the sites in the center of the lattice are doubly occupied.
Experimentally, this is achieved by using an additional trapping potential
to confine the fermions to a central region of the optical lattice
\citep{Schneider2012}. At time $t=0$ the trapping potential is switched
off and the fermions expand into the empty lattice sites, as depicted
in figure~\ref{fig: Schematic}(a). 

The time evolution of the average fermion density and the average
doublon density has already been studied for this expansion protocol,
by numerical simulations for 1D lattices \citep{Kajala2011}. For
large onsite interactions strengths $U/J\gtrsim4$ two wave fronts
with different velocities are visible in the time-dependent density
profile, while there is a single wave front for small onsite interaction
strengths. It turns out that the expansion can be basically understood
as a mixture of propagating single fermions and doublons, see also
\citep{Langer2012}. For small onsite interaction strengths the initial
state quickly decays into single fermions, which move ballistically
through the lattice. Due to the cosine dispersion relation of a single
particle in the lattice, $\epsilon_{k}=-2J\cos(k)$ with wavenumber
$k\in(-\pi,\pi]$ , its maximal velocity is given by $|v_{max}|=2J$.
This fact leads to light cones in the density distribution, as indicated
in figure~\ref{fig: Schematic}(a). For large onsite interaction
$U/J\gtrsim4$, only a small fraction of doublons decay into fermions,
which move away ballistically. As the effective tunneling amplitude
of a doublon is $2J^{2}/U$, they initially remain in the central
region and then move through the lattice on a much larger time scale.

However, the propagation of the single fermions is highly correlated,
as they are always created as spin-singlet pair when a doublon decays.
As sketched in figures~\ref{fig: Schematic}(b) and (c), the two
particles of such a spin-entangled pair may be either detected on
the same side of the initial cluster or on different sides. 

We briefly point out that the dynamics of the spin-entanglement as
well as that of the density-density correlation is invariant under
the change of the sign in the interaction strength $U.$ This is a
direct consequence of a transformation property of the (spin- or density-)
correlators and of the initial state under time reversal and $\pi$-boost
(translation of all momenta by $\pi$). This dynamical $U\mapsto-U$
symmetry in the Hubbard model is discussed in more detail in \citep{Schneider2012,Deuchert2012}.

\subsection{Numerical simulation}

For the numerical evaluation of the nonequilibrium time evolution,
we employ the time-dependent density-matrix renormalization group
(tDMRG) method \citep{Vidal2004,Daley2004,White2004,Schmitteckert2004,Schollwoeck2011}.
The initial state is a cluster consisting of 10 doublons, which are
located in the center of a lattice of size $L=100$ with open boundary
conditions. Our tDMRG simulation uses a Krylov subspace method with
time step $J\delta t=0.125$ and the discarded weight is set to either
$10^{-5}$ or $10^{-6}$, depending on the interaction strength. For
all evolution times shown in the figures, the density remains negligible
at the boundaries of the lattice. We also verified that the features
presented here does not depend on the precise position of the cluster
in the lattice nor change when the truncation error is modified within
the given range. 

For non-interacting particles, we have used the exact expressions
for the time-dependent correlation functions and concurrence, as given
in appendix A.

\section{Density-density correlation\label{sec:Density-density-correlation}}

Before turning to the spin-entanglement (in the subsequent section),
we will first study the density-density correlation function $D_{ij}(t)=\langle\hat{n}_{i}(t)\hat{n}_{j}(t)\rangle-\langle\hat{n}_{i}(t)\rangle\langle\hat{n}_{j}(t)\rangle$,
with $\hat{n}_{i}=\hat{n}_{i,\uparrow}+\hat{n}_{i,\downarrow}$. For
the expansion, sketched in figure~\ref{fig: Schematic}, the density
at different lattice sites is expected to be correlated for several
reasons, in particular since the fermions are always created in pairs
out of doublons.

Note that the single-site detection of density correlations (more
precisely the parity correlations) has been very recently used to
study the quasi-particle propagation in a commensurate ultracold bosonic
gas after an interaction quench \citep{Cheneau2012,Barmettler2012}.
It has also been employed to study the role of onsite interaction
in a bosonic two-body quantum walk, experimentally realized in a nonlinear
optical waveguide lattice \citep{Lahini2012}.

\begin{figure}
\begin{raggedleft}
\includegraphics[scale=0.28]{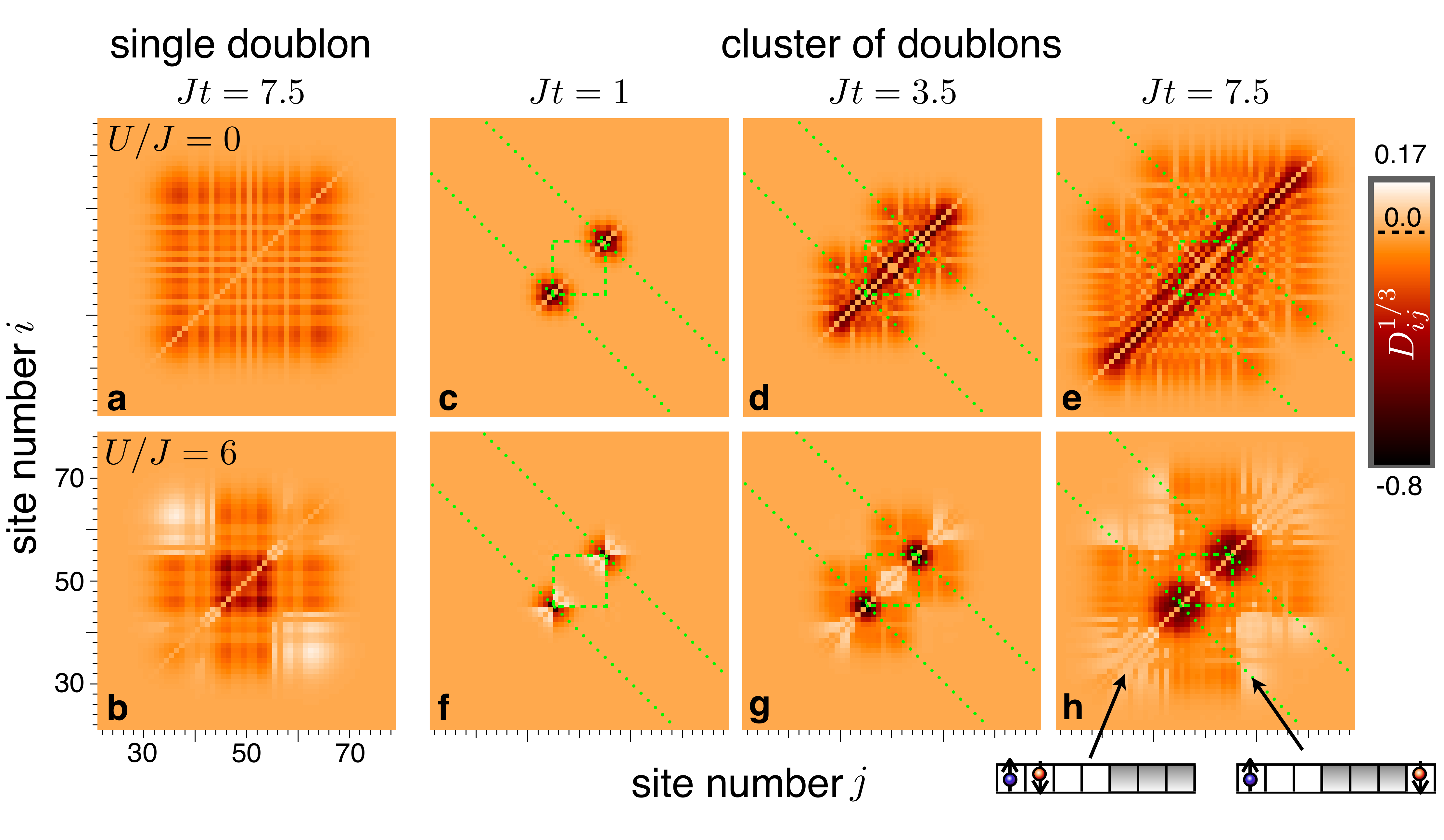}
\par\end{raggedleft}

\caption{Off-diagonal density correlations $D_{ij\neq i}(t)=\langle\hat{n}_{i}(t)\hat{n}_{j\neq i}(t)\rangle-\langle\hat{n}_{i}(t)\rangle\langle\hat{n}_{j\neq i}(t)\rangle$
for the expansion from an initial state consisting of a single doublon
(a,b) and a cluster of ten doublons (c-h). For better visibility we
display the values of $D_{ij}^{1/3}$ as a colour scale and set $D_{ii}$
to zero. (c,f) Initially all fermions but the outermost ones are Pauli-blocked.
This leads to non-vanishing density-density correlations only close
to the edge of the initial configuration for small evolution times.
(a,c-e) For vanishing onsite interaction $U$, the density-density
correlation equals the two-site spin-$z$ correlation and is always
nonpositive, cf. equations~(\ref{eq:spin-spin correlation noninteracting case})
and (\ref{eq:relation of spin and density correlation noninteracting case}).
(b,f-h) In the interacting case, the density-density correlation can
be positive. (b) For a single doublon, $D_{ij\neq i}(t)>0$ only for
$i$ and $j$ at different sides of the initially occupied lattice
site. Thus, the doublon decays primarily into fermions moving in opposite
directions. (f-h) For a cluster of doublons, positive density correlations
are also found for lattice sites $i$ and $j$ that are both located
to the left or to the right of the initial cluster position, or within
the initial cluster (indicated by the dashed square). \label{fig:density-density-fluctuation}}

\end{figure}

As a point of reference, we first consider the density-density correlations
for the initial state consisting of a single doublon. The fluctuation
in the density at a lattice site $i$, $D_{ii}(t)$, is the variance
of the occupation number $\hat{n}_{i}$. For the expansion from a
doublon, it is maximal for those lattice sites located at the edge
of the single fermion light cone (doublon light cone) in the noninteracting
(strongly interacting) case. In the following we focus on the off-diagonal
density-density correlations $D_{ij\neq i}(t)$, which are shown in
figures~\ref{fig:density-density-fluctuation}(a) and (b). Most importantly,
we find that the onsite interaction $U$ leads to a positive correlation
of those fermions propagating with almost maximal velocity $|2J|$
in opposite directions, see figure~\ref{fig:density-density-fluctuation}(b).
On the other hand, the density-density correlation assumes large negative
values between those lattice sites in the center that are most likely
occupied by the doublon (and not by single fermions) {[}central square
region in figure~\ref{fig:density-density-fluctuation}(b){]}. In
contrast, the off-diagonal density-density correlations are always
nonpositive for clusters of noninteracting fermions, as detailed in
appendix A, see also figures~\ref{fig:density-density-fluctuation}(a)
and (c)-(e).

The bunching effect in the interacting case can be understood by writing
the motion of the two fermions in relative and center of mass coordinates,
$r$ and $R$, respectively (see appendix B for details of the discussion).
Note that we assume an infinite lattice for this argument, which is
compatible with the simulation as the boundary conditions do not play
a role for the evolution times considered here. The center of mass
motion is a plane wave with total wavenumber $K=(k_{1}+k_{2})\,\mbox{mod}\,2\pi$,
where $k_{1}$ and $k_{2}$ are the asymptotic wave numbers of the
single fermions. The relative motion is described by a $K$-dependent
Hamiltonian. For $U\neq0$, the eigenstates of this Hamiltonian are
one bound state and scattering states. The probability for the doublon
to decay into one specific scattering state with wavefunction $\psi_{K,k}(r)$
{[}where $k=(k_{1}-k_{2})/2$ is the relative wavenumber{]} is given
by the modulus squared of that wavefunction's probability amplitude
at $r=0$: $|\psi_{K,k}(0)|^{2}$. This decay probability is, up to
an overall normalization constant {[}see equation~(\ref{eq:squared overlap of doublon and scattering states}){]}:\begin{equation}
|\psi_{K,k}(0)|^{2}\propto\left[1+U^{2}/\left(16J^{2}\cos^{2}(K/2)\sin^{2}(k)\right)\right]^{-1}.\end{equation}
For small onsite interactions $|U/4J|\ll1$, the decay probability
is almost the same for the different scattering states, except for
$K\approx\pi$ or $k\approx0$ where it drops to zero. In the strongly
interacting case, it exhibits a pronounced maximum for $K=0$ and
$k=\pm\pi/2$, that is $k_{1}=\pm\pi/2$ and $k_{2}=\mp\pi/2$. In
other words, for large onsite interaction the doublon decays primarily
into two fermions moving in opposite directions with velocity close
to $|2J|$.

Next we study the density correlations for the expansion from a cluster
of several doublons into an empty lattice. The results are displayed
in figures~\ref{fig:density-density-fluctuation}(c)-(h). Just as
for the single doublon, the off-diagonal density correlation has regions
of positive values in the presence of onsite interaction. However,
in contrast to the case of a single doublon, the existence of other
doublons now leads to positive correlations also at lattice sites
located on the same side of the initial cluster, see figures~\ref{fig:density-density-fluctuation}(f)-(h).
Positive correlations between sites on different sides of the cluster
are only observed for evolution times larger than the time a hole
takes to propagate through the cluster, cf. figure~\ref{fig:density-density-fluctuation}(h).

The results suggest that the initially Pauli blocked core of the cluster
leads to an enhanced decay of the outermost doublons into single fermions
moving away in the same direction. However, the alternative case,
where the edge doublon decays into a fermion and a hole moving into
the opposite direction close to the maximal velocity $|2J|$ also
leads to positive correlations. Further simulations show that positive
density correlations between lattice sites on the same side of the
initial cluster positions are found for clusters of about four and
more doublons and $U/J\gtrsim2$. Note that as the single fermions
are created by the decay of an edge doublon, the situation is different
from a free single fermion or fermionic wave packet approaching a
cluster of doublons. In the latter situations, the fermion is almost
perfectly transmitted through the cluster (band insulating state)
in the limit of large onsite interaction \citep{Muth2012,Jin2011}. 

In the next section, we discuss the spin-entanglement. This will reveal
that the positive density correlations indeed stem from singlet pairs,
which become delocalized by the decay of a doublon.

\section{Spin-entanglement between different lattice sites within the expanding
cloud\label{sec:Spin-entanglement}}

In this section, we discuss the entanglement between fermions located
at different lattice sites by means of the concurrence \citep{Wootters1998}.
First, the relation between the concurrence and the spin-spin correlations
is established. Then we use tDMRG simulations to find the regions
where fermions are likely and unlikely to be entangled during the
expansion. We examine the role of the onsite interaction and the core
of the cluster on the entanglement dynamics.

\subsection{Reduced density matrix and concurrence of two fermions\label{sub: reduced density matrix and concurrence}}

In this paragraph we derive the reduced density matrix and the concurrence
for fermions located at different lattice sites within the expanding
cloud (see \citep{Ramsak2009b} for a related discussion in the context
of solid state physics). Given the full many-body wavefunction $\ket{\Psi(t)}$,
the reduced density matrix for the lattice sites $i$ and $j$ is
$\hat{\rho}_{ij}(t)=\mbox{Tr}_{{\rm sites}\neq i,j}\{\ket{\Psi(t)}\bra{\Psi(t)}\}$,
where all other lattice sites have been traced out. Whenever two fermions
are situated at the same lattice site they form a spin-singlet pair
as their spatial degrees of freedom are symmetric under particle exchange.
Thus, entanglement in the spin degree of freedom between fermions
at two different lattice sites can only occur if each lattice site
is occupied by a single fermion. Projecting $\hat{\rho}_{ij}(t)$
onto those states yields

\begin{equation}
\hat{\rho}_{ij}^{s}(t)=\frac{1}{\mbox{Tr}\{\hat{\rho}_{ij}(t)\,\hat{n}_{i}^{s}\hat{n}_{j}^{s}\}}\hat{n}_{j}^{s}\hat{n}_{i}^{s}\,\hat{\rho}_{ij}(t)\,\hat{n}_{i}^{s}\hat{n}_{j}^{s}.\end{equation}
Here, $\hat{n}_{i}^{s}=\hat{n}_{i,\uparrow}+\hat{n}_{i,\downarrow}-2\hat{n}_{i,\uparrow}\hat{n}_{i,\downarrow}$
is the single fermion number operator at site $i$, which projects
onto a subspace with exactly one fermion on that site. The normalization
factor in the denominator, $\mbox{Tr}\{\hat{\rho}_{ij}(t)\hat{n}_{i}^{s}\hat{n}_{j}^{s}\}=\langle\hat{n}_{i}^{s}(t)\hat{n}_{j}^{s}(t)\rangle$,
is the probability of finding at time $t$ a single fermion at each
of the lattice sites $i$ and $j$. The state described by $\hat{\rho}_{ij}^{s}(t)$
is the state obtained after a successful projective measurement. The
reduced density matrix $\hat{\rho}_{ij}^{s}(t)$ is equivalent to
a two-qubit density matrix, which can be expressed in the form $\hat{\rho}=\sum_{\alpha,\beta=0}^{3}\lambda^{\alpha\beta}\,\hat{\sigma}_{(1)}^{\alpha}\otimes\hat{\sigma}_{(2)}^{\beta}$,
where $\hat{\sigma}^{1,2,3}$ are the Pauli matrices, $\hat{\sigma}^{0}=\mathbf{1}$,
and the factors $\lambda^{\alpha\beta}$ are determined by the correlation
functions $\lambda^{\alpha\beta}=\frac{1}{4}\langle\hat{\sigma}_{(1)}^{\alpha}\hat{\sigma}_{(2)}^{\beta}\rangle$
\citep{Fano1957}. Consequently, the reduced density matrix of single
fermions at lattice sites $i$ and $j$ can be written as

\begin{equation}
\hat{\rho}_{ij}^{s}(t)=\frac{1}{\langle\hat{n}_{i}^{s}(t)\hat{n}_{j}^{s}(t)\rangle}\sum_{\alpha,\beta=0}^{3}\langle\hat{S}_{i}^{\alpha}(t)\hat{S}_{j}^{\beta}(t)\rangle\,\hat{\sigma}_{i}^{\alpha}\otimes\hat{\sigma}_{j}^{\beta},\label{eq:reduced density matrix general form}\end{equation}
where $\hat{S}_{i}^{1,2,3}=\frac{1}{2}\sum_{a,b=\uparrow,\downarrow}\hat{c}_{i,a}^{\dagger}(\hat{\sigma}^{1,2,3})_{\ b}^{a}\hat{c}_{i,b}$
is the $x$-, $y$-, and $z$-component of the spin operator and $\hat{S}_{i}^{0}$
is, for compactness, defined as half the single fermion number operator,
$\hat{S}_{i}^{0}:=\frac{1}{2}\hat{n}_{i}^{s}$. Note that $\langle\hat{S}_{i}^{\alpha}(t)\hat{S}_{j}^{\beta}(t)\rangle$
is calculated using the full (unprojected) wavefunction since states
with vacancies or doublons at site $i$ or $j$ do not contribute
to the expectation value. 

Symmetries of the initial state and the Hamiltonian can simplify the
form of the reduced density matrix, such that only a few correlation
functions are needed to determine $\hat{\rho}_{ij}^{s}(t)$. As detailed
now, the reduced density matrix $\hat{\rho}_{ij}^{s}(t)$ depends
only on $\langle\hat{S}_{i}^{z}(t)\hat{S}_{j}^{z}(t)\rangle/\langle\hat{n}_{i}^{s}(t)\hat{n}_{j}^{s}(t)\rangle$
for the cluster initial state shown in figure~\ref{fig: Schematic}.
The Hubbard Hamiltonian (\ref{eq:FH Hamiltonian}) preserves the spin-dependent
particle number, i.e., $[\hat{\mathcal{H}},\hat{N}_{\uparrow,\downarrow}]=0$,
$\hat{N}_{\uparrow,\downarrow}=\sum_{i=1}^{L}\hat{n}_{\uparrow,\downarrow}$.
Given an initial state with fixed number of spin-up and spin-down
fermions, which is the usually case in cold atom experiments, the
time-dependent expectation values of operators that do change the
spin-dependent particle number vanish. This yields $\langle\hat{n}_{i}^{s}(t)\hat{S}_{j}^{x,y}(t)\rangle=0$,
$\langle\hat{S}_{i}^{x,y}(t)\hat{S}_{j}^{z}(t)\rangle=0$, and $\langle\hat{S}_{i}^{x}(t)\hat{S}_{j}^{x}(t)\rangle-\langle\hat{S}_{i}^{y}(t)\hat{S}_{j}^{y}(t)\rangle\pm i[\langle\hat{S}_{i}^{x}(t)\hat{S}_{j}^{y}(t)\rangle+\langle\hat{S}_{i}^{y}(t)\hat{S}_{j}^{x}(t)\rangle]=0$.
The latter condition comes from creating two spin-down (spin-up) fermions
while destroying two spin-up (spin-down) fermions at lattice sites
$i$ and $j$ and can also be written as $\langle\hat{S}_{i}^{x}(t)\hat{S}_{j}^{x}(t)\rangle=\langle\hat{S}_{i}^{y}(t)\hat{S}_{j}^{y}(t)\rangle$
and $\langle\hat{S}_{i}^{x}(t)\hat{S}_{j}^{y}(t)\rangle=-\langle\hat{S}_{i}^{y}(t)\hat{S}_{j}^{x}(t)\rangle$.
Moreover, the Hamiltonian (\ref{eq:FH Hamiltonian}) is fully rotationally
invariant, $[\hat{\mathcal{H}},\sum_{i=1}^{L}\hat{S}_{i}^{x,y,z}]=0$.
If the initial state is rotationally invariant, for instance a cluster
of doublons, then the many-body state remains SU(2) spin symmetric
during the time evolution. It follows that $\langle\hat{S}_{i}^{z}(t)\hat{S}_{j}^{z}(t)\rangle=\langle\hat{S}_{i}^{x}(t)\hat{S}_{j}^{x}(t)\rangle=\langle\hat{S}_{i}^{y}(t)\hat{S}_{j}^{y}(t)\rangle$,
$\langle\hat{n}_{i}^{s}(t)\hat{S}_{j}^{z}(t)\rangle=\langle\hat{n}_{i}^{s}(t)\hat{S}_{j}^{x,y}(t)\rangle=0$,
and $\langle\hat{S}_{i}^{x}(t)\hat{S}_{j}^{y}(t)\rangle=\langle\hat{S}_{i}^{x,y}(t)\hat{S}_{j}^{z}(t)\rangle=0$.
In summary, the reduced density matrix for the expansion from a cluster
of doublons reads \begin{eqnarray}
\hat{\rho}_{ij}^{s}(t) & = & \frac{1}{4}\cdot\mathbf{1}+\frac{\langle\hat{S}_{i}^{z}(t)\hat{S}_{j}^{z}(t)\rangle}{\langle\hat{n}_{i}^{s}(t)\hat{n}_{j}^{s}(t)\rangle}\left[\hat{\sigma}_{i}^{x}\otimes\hat{\sigma}_{j}^{x}+\hat{\sigma}_{i}^{y}\otimes\hat{\sigma}_{j}^{y}+\hat{\sigma}_{i}^{z}\otimes\hat{\sigma}_{j}^{z}\right]\nonumber \\
 & = & \bigl(\frac{1}{4}+\frac{\langle\hat{S}_{i}^{z}(t)\hat{S}_{j}^{z}(t)\rangle}{\langle\hat{n}_{i}^{s}(t)\hat{n}_{j}^{s}(t)\rangle}\bigr)\left[\ket{T_{ij}^{1}}\bra{T_{ij}^{1}}+\ket{T_{ij}^{0}}\bra{T_{ij}^{0}}+\ket{T_{ij}^{-1}}\bra{T_{ij}^{-1}}\right]\nonumber \\
 &  & +\bigl(\frac{1}{4}-3\frac{\langle\hat{S}_{i}^{z}(t)\hat{S}_{j}^{z}(t)\rangle}{\langle\hat{n}_{i}^{s}(t)\hat{n}_{j}^{s}(t)\rangle}\bigr)\ket{S_{ij}}\bra{S_{ij}},\label{eq:reduced density matrix reduced form}\end{eqnarray}
where $\ket{S_{ij}}=\frac{1}{\sqrt{2}}(\ket{\uparrow_{i},\downarrow_{j}}-\ket{\downarrow_{i},\uparrow_{j}})$
is the singlet state and $\ket{T_{ij}^{m}}$ is the triplet state
with the $m$ denoting the spin projection in $z$-direction. 

Instead of $\langle\hat{S}_{i}^{z}(t)\hat{S}_{j}^{z}(t)\rangle$ one
could in principle evaluate the spin-spin correlation in any other
direction. However, for cold atomic gases in optical lattices the
correlation function $\langle\hat{S}_{i}^{z}(t)\hat{S}_{j}^{z}(t)\rangle=\frac{1}{4}\langle[\hat{n}_{i,\uparrow}(t)-\hat{n}_{i,\downarrow}(t)][\hat{n}_{j,\uparrow}(t)-\hat{n}_{j,\downarrow}(t)]\rangle$
seems to be experimentally most realistic to access as it could be
obtained from snapshots of the spin-dependent single-site detection
of the particle number. 

The spin-entanglement between single fermions can be derived from
the reduced density matrix. In this work, we use the concurrence $C(\hat{\rho})$
\citep{Wootters1998} to quantify the entanglement. Given the time-reversed
density matrix $\hat{\tilde{\rho}}=\sigma^{y}\otimes\sigma^{y}\hat{\rho}^{*}\sigma^{y}\otimes\sigma^{y}$,
with the complex conjugation $\hat{\rho}^{*}$ taken in the standard
basis $\{\ket{\uparrow\uparrow},\ket{\uparrow\downarrow},\ket{\downarrow\uparrow},\ket{\downarrow\downarrow}\}$,
the concurrence is defined by $C(\hat{\rho})=\mbox{max}\{0,\sqrt{\lambda_{1}}-\sqrt{\lambda_{2}}-\sqrt{\lambda_{3}}-\sqrt{\lambda_{4}}\}$,
where the $\lambda_{i}$'s are the eigenvalues of $\hat{\rho}\hat{\tilde{\rho}}$
in descending order. In our case, the concurrence of the reduced density
matrix (\ref{eq:reduced density matrix reduced form}) is given by

\begin{equation}
C_{i,j}(t)=\mbox{max}\left\{ 0,-\frac{1}{2}-6\frac{\langle\hat{S}_{i}^{z}(t)\hat{S}_{j}^{z}(t)\rangle}{\langle\hat{n}_{i}^{s}(t)\hat{n}_{j}^{s}(t)\rangle}\right\} .\label{eq:Concurrence}\end{equation}
Spin-spin correlations $\langle\hat{S}_{i}^{z}(t)\hat{S}_{j}^{z}(t)\rangle/\langle\hat{n}_{i}^{s}(t)\hat{n}_{j}^{s}(t)\rangle\geq-1/12$
result in a vanishing concurrence. The concurrence approaches $1$
as $\langle\hat{S}_{i}^{z}(t)\hat{S}_{j}^{z}(t)\rangle/\langle\hat{n}_{i}^{s}(t)\hat{n}_{j}^{s}(t)\rangle\searrow-1/4$,
i.e., when the two fermions detected at lattice sites $i$ and $j$
always have opposite spin.

The probability that single fermions at lattice sites $i$ and $j$
form a spin-singlet pair can be directly read off the reduced density
matrix (\ref{eq:reduced density matrix reduced form}):

\begin{equation}
P_{ij}^{Singlet}(t)=\mbox{Tr}\left[\hat{\rho}_{ij}^{s}(t)\ket{\mathcal{S}_{ij}}\bra{\mathcal{S}_{ij}}\right]=\frac{1}{4}-3\frac{\langle\hat{S}_{i}^{z}(t)\hat{S}_{j}^{z}(t)\rangle}{\langle\hat{n}_{i}^{s}(t)\hat{n}_{j}^{s}(t)\rangle}.\label{eq:Singlet probability}\end{equation}
Each of the spin triplet states is measured with the probability $1/4+\langle\hat{S}_{i}^{z}(t)\hat{S}_{j}^{z}(t)\rangle/\langle\hat{n}_{i}^{s}(t)\hat{n}_{j}^{s}(t)\rangle$.

\subsection{Time evolution of the spin-entanglement within the expanding cloud}

\begin{figure}
\begin{raggedleft}
\includegraphics[scale=0.28]{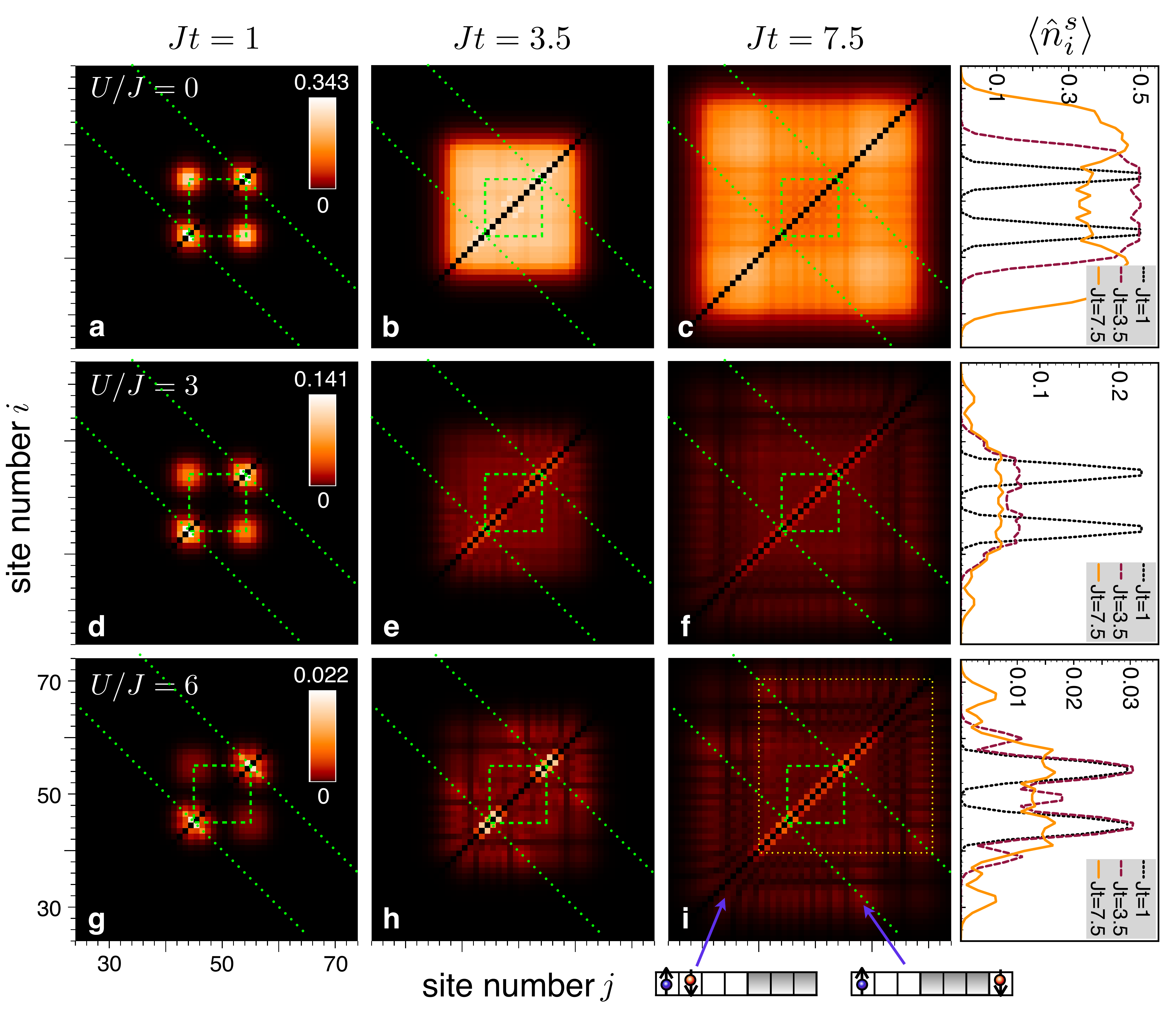}
\par\end{raggedleft}

\caption{Density-density correlation of single fermions $\langle\hat{n}_{i}^{s}(t)\hat{n}_{j\neq i}^{s}(t)\rangle$
(i.e. excluding doublons) for the expansion from an initial cluster
of doublons in one dimension (the initial position of the ten doublons
is indicated by the dashed square). (a-c) Without onsite interactions,
$U/J=0$, the fermions move ballistically through the lattice. This
is reflected by the fourfold symmetry of the correlation matrix shown
here. (d-i) When increasing $U/J$, fermions are created only rarely
by the decay of a doublon at the edge of the cluster. These two fermions
move within the same light cone, cf. figure~\ref{fig: Schematic}.
The light cone leads to a square shape of the correlation function
having its center at the edge of the initial cluster, shown by the
dotted square in panel (i). The dotted lines indicate pairs of coordinates
corresponding to fermions emitted with opposite velocities by a doublon
at the edge of the cloud. Moreover, an increased correlation between
nearest neighbor lattice sites is found for larger evolution times,
see (f) and (i). \label{fig:Density-density-correlation-unpaired-fermions}}

\end{figure}

Spin-entanglement between different lattice sites, which we denote
by $i$ and $j$, requires that both sites are singly occupied as
discussed in section~\ref{sub: reduced density matrix and concurrence}.
The probability for this is $\langle\hat{n}_{i}^{s}(t)\hat{n}_{j}^{s}(t)\rangle$,
with the single fermion number operator already defined above ($\hat{n}_{i}^{s}=\hat{n}_{i,\uparrow}+\hat{n}_{i,\downarrow}-2\hat{n}_{i,\uparrow}\hat{n}_{i,\downarrow}$).
Figure~\ref{fig:Density-density-correlation-unpaired-fermions} shows
the numerical results for $\langle\hat{n}_{i}^{s}(t)\hat{n}_{j\neq i}^{s}(t)\rangle$
for different evolution times $Jt$ and onsite interaction strengths
$U/J$. 

Shortly after the quench, single fermions are created at the edges
of the cluster, see figures~\ref{fig:Density-density-correlation-unpaired-fermions}(a),(d),
and (g). For the noninteracting system, the fermions escape the cluster
successively (i.e., starting at the edges, and finally from the center
of the cluster). They move ballistically and the correlation function
displays a fourfold symmetric structure, figures~\ref{fig:Density-density-correlation-unpaired-fermions}(b)
and (c). In the interacting case, in contrast, the decay of a doublon
into fermions is heavily suppressed. This results in a correlation
function $\langle\hat{n}_{i}^{s}(t)\hat{n}_{j}^{s}(t)\rangle$ that
has its main contributions within two square regions given by the
light cones of fermions emitted by the outermost doublons of the cluster,
see figures~\ref{fig:Density-density-correlation-unpaired-fermions}(f)
and (i). Within the light cones, single fermions are more likely to
be found at lattice sites corresponding to a motion with almost maximal
velocity $|v_{max}|=2J$ into opposite direction. The density correlation
between single fermions attains its largest values for nearest-neighbor
lattice sites within the cloud, cf. figures~\ref{fig:Density-density-correlation-unpaired-fermions}(e),(f),(h),
and (i). This may be due to virtual transitions between two configurations:
either a doublon with a neighboring vacancy, or a state of two adjacent
single fermions. Note that the mean total number of single fermions,
$\sum_{i=1}^{L}\langle\hat{n}_{i}^{s}\rangle$, approaches within
a few $Jt$ an almost constant value, cf. area under the curves for
$Jt=3.5$,$7.5$ and $U/J=3,6$ in the last column of figure~\ref{fig:Density-density-correlation-unpaired-fermions}.
That means only the edges of the cluster evaporate for large interactions,
releasing a finite number of single fermions. Moreover, $\langle\hat{n}_{i}^{s}(t)\rangle$
becomes relatively flat in the center of the cloud for larger evolution
times $Jt.$ 

\begin{figure}
\begin{raggedleft}
\includegraphics[scale=0.28]{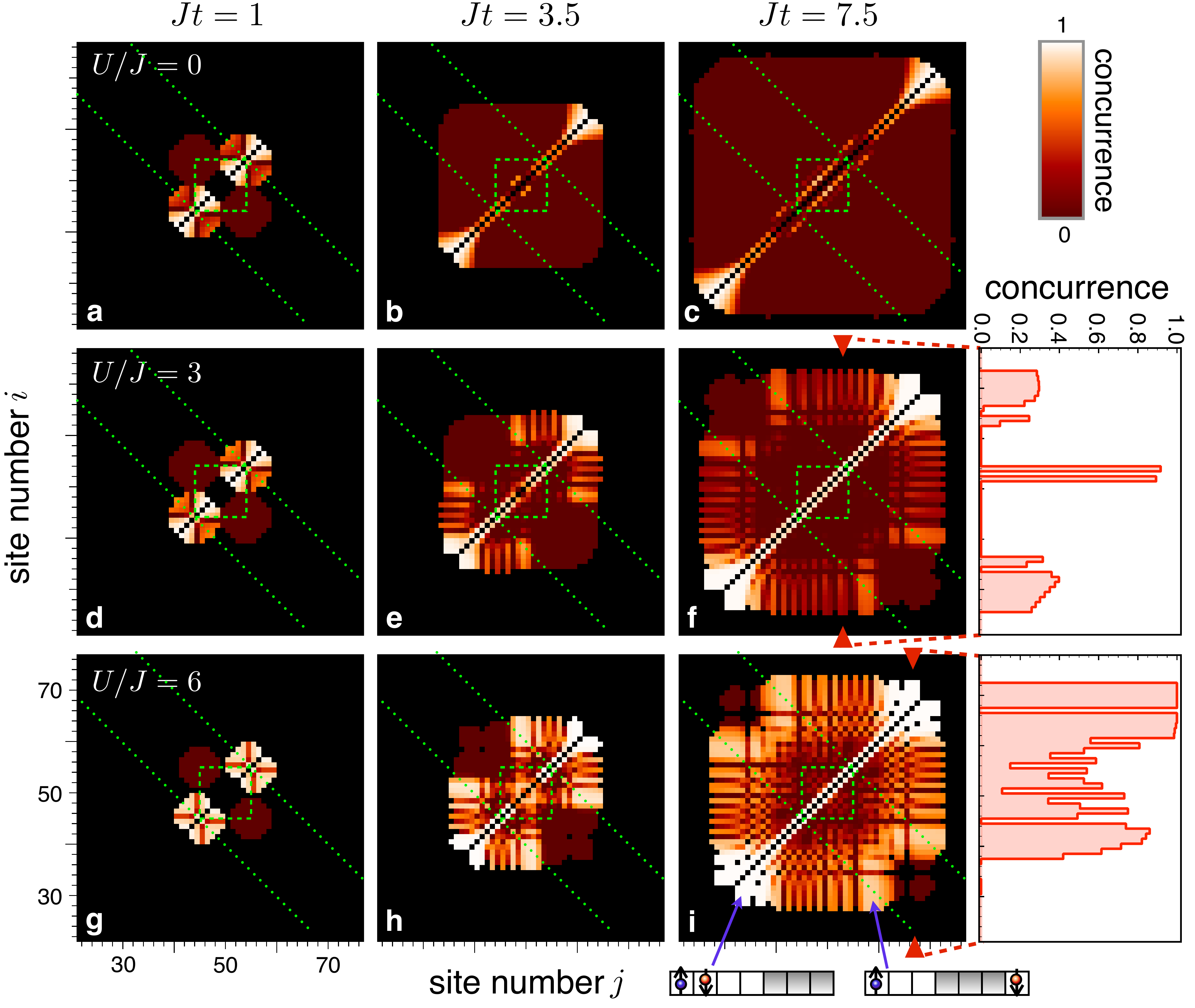}
\par\end{raggedleft}

\caption{Concurrence $C_{i,j}(t)$ of two (single) fermions located at lattices
sites $i$ and $j$ after different expansion times $Jt$, see equation~(\ref{eq:Concurrence}).
Initially, ten doublons are located at the sites indicated by the
dashed square. We emphasize that dark red indicates strictly zero
concurrence (also see the cuts displayed to the right). The black
colour code is used whenever the probability of finding fermions at
sites $i$ and $j$ is too small, $\langle\hat{n}_{i}^{s}(t)\hat{n}_{j}^{s}(t)\rangle<10^{-5}$.
In those cases, the concurrence is not computed as it would become
susceptible to numerical inaccuracies. (a,d,g) For short evolution
times, here $Jt=1$, single fermions are mainly created at the edge
of the cluster since the central fermions are initially Pauli blocked.
Fermions close to the same edge of the cloud are likely to be entangled
as they are likely to originate from the same doublon. Fermions at
opposite edges are not entangled since they can not have been emitted
from the same doublon. (b,c) At larger evolution times and without
onsite interactions, the concurrence is 0 for most sites $i$ and
$j$. However, it is close to 1 when two fermions are located at the
outermost part of the cloud. (e,f,h,i) By increasing the onsite interaction
$U/J$, entanglement of fermions across the cluster becomes possible.
In this case, the concurrence is highest when the fermions result
from the decay of a doublon at the edge and escape with opposite velocity,
indicated by the dotted line {[}see also cut through panel (i){]}.
Moreover, the concurrence of fermions at neighboring sites becomes
almost 1. Within the central region, approximately given by the overlap
of the light cones shown in figure~\ref{fig: Schematic}, the concurrence
remains relatively small. \label{fig:Concurrence-unpaired-fermions}}

\end{figure}

Let us now consider the spatial distribution of spin-entangled fermions
during the expansion. For this purpose, the concurrence between two
fermions at different lattice sites is numerically evaluated using
equation~(\ref{eq:Concurrence}). Figure~\ref{fig:Concurrence-unpaired-fermions}
shows the concurrence for any pair of lattice sites $i$ and $j$,
at different times $Jt$ and onsite interaction strengths $U/J$. 

For the expansion of noninteracting fermions, figures~\ref{fig:Concurrence-unpaired-fermions}(a)-(c),
the concurrence is finite only for nearby lattice sites. It is almost
1 within the outermost wings of the expanding cloud. This can be physically
understood the following way: Due to the Pauli principle, fermions
with the same spin become spatially antibunched during the expansion.
Thus, fermions at neighboring sites are more likely to have opposite
spin, cf. figures~\ref{fig:density-density-fluctuation}(c)-(e).
In addition, the outermost region of the cloud lies only within the
light cones of the doublons close to the edge of the initial cluster.
Two fermions detected in this region are almost certainly emitted
by the same edge doublon and in consequence have a high probability
to be spin-entangled.

When increasing the onsite interaction $U$, spin-entangled pairs
are formed on remote lattice sites, too, cf. figures~\ref{fig:Concurrence-unpaired-fermions}(d)-(i).
Indeed, spin-entanglement is found between lattice sites within and
outside the initial cluster position, figures~\ref{fig:Concurrence-unpaired-fermions}(e)
and (h), as well as on different sides of the initial cluster position,
figures~\ref{fig:Concurrence-unpaired-fermions}(f) and (i). 

The figures are a fingerprint of the creation of a counter propagating
hole and single fermion by the decay of an edge doublon. When the
hole moves through the cluster, the spin-entanglement with the single
fermion outside the cluster is swapped sequentially from one fermion
to the next in the cluster. In this way fermions become entangled
that have never been on the same lattice site and have never directly
interacted with each other. At the end of this process, a spin-singlet
pair is created with a fermion at each side of the cluster. The concurrence
for two fermions on different sides of the initial cluster position
increases with the interaction strength, compare figures~\ref{fig:Concurrence-unpaired-fermions}(f)
and \ref{fig:Concurrence-unpaired-fermions}(i). This can be understood
by the suppression of the decay probability of doublons with increasing
interaction strengths. For larger interaction strength, a hole is
more likely to cross the cluster without being disturbed by another
hole.

For nearest-neighbor sites, which have a relatively large probability
to be simultaneously singly occupied {[}see figures~\ref{fig:Density-density-correlation-unpaired-fermions}(f)
and (i){]}, we observe a concurrence close to 1. This implies the
creation of vacant lattice sites within the cloud during the expansion.
The singlet pairs on nearest-neighbor sites come from virtual transitions
between a doublon with neighboring vacancy and a state of two single
fermions. We verified this behaviour in addition by numerically creating
an ensemble of snapshots for the distribution of fermions as described
in \citep{Kessler2012}.

We emphasize that for inhomogeneous systems, such as the one discussed
in this article, the spatially resolved measurement of two-point correlations
provides more information about the dynamics than structure factors,
such as $S_{k}\propto\sum_{lm}e^{-ik(l-m)}\langle\hat{S}_{l}^{z}\hat{S}_{m}^{z}\rangle$.
While the latter could be used to determine the average spin-spin
correlation of fermions at fixed distance, it contains no information
where these spin-correlated pairs are located in the cloud.

\subsection{Summed concurrences\label{sub:Summed-concurrences}}

Above, we examined the spin-entanglement between two lattice sites
for fixed time points. In this subsection we aim to quantify the spin-entanglement
for entire regions in the lattice. In particular, we address the questions:
Are there sites which share more spin-entanglement with the rest than
other lattice sites? How does the spin-entanglement in different regions
built up as function of time? Which locations are most entangled in
the weakly and strongly interacting case? How does the size of the
cluster affect these results?

In the following we discuss the amount of pairwise spin-entanglement
of a lattice site or a region in the lattice in terms of the \emph{summed
concurrence}. We define it as the sum over the concurrences, $C_{i,j}(t)$,
which are weighted by the probability of detecting a single fermion
at both lattice sites, $\langle\hat{n}_{i}^{s}(t)\hat{n}_{j}^{s}(t)\rangle$.
The weights are introduced to accommodate for the possibility of vacant
or doubly occupied lattice sites, in contrast to the summed concurrence
used in spin systems \citep{Amico2004}. For a system consisting of
spin-singlet pairs whose wavefunctions do not overlap, i.e., $C_{i,j}(t)$
is either zero or one for all sites $i$ and $j$, the summed concurrence
equals the average number of delocalized spin-singlet pairs. Note
that the summed concurrence is by no means a measure for the total
entanglement of the system. It neither includes multipartite entanglement
nor entanglement in the occupation numbers. For a detailed discussion
on entanglement in many-body systems we refer to \citep{Amico2008},
see also \citep{Vollbrecht2007,Cramer2011b,Levine2011,Daley2012}
for recent proposals on detecting entanglement in cold atom systems.

\begin{figure}
\begin{raggedleft}
\includegraphics[scale=0.28]{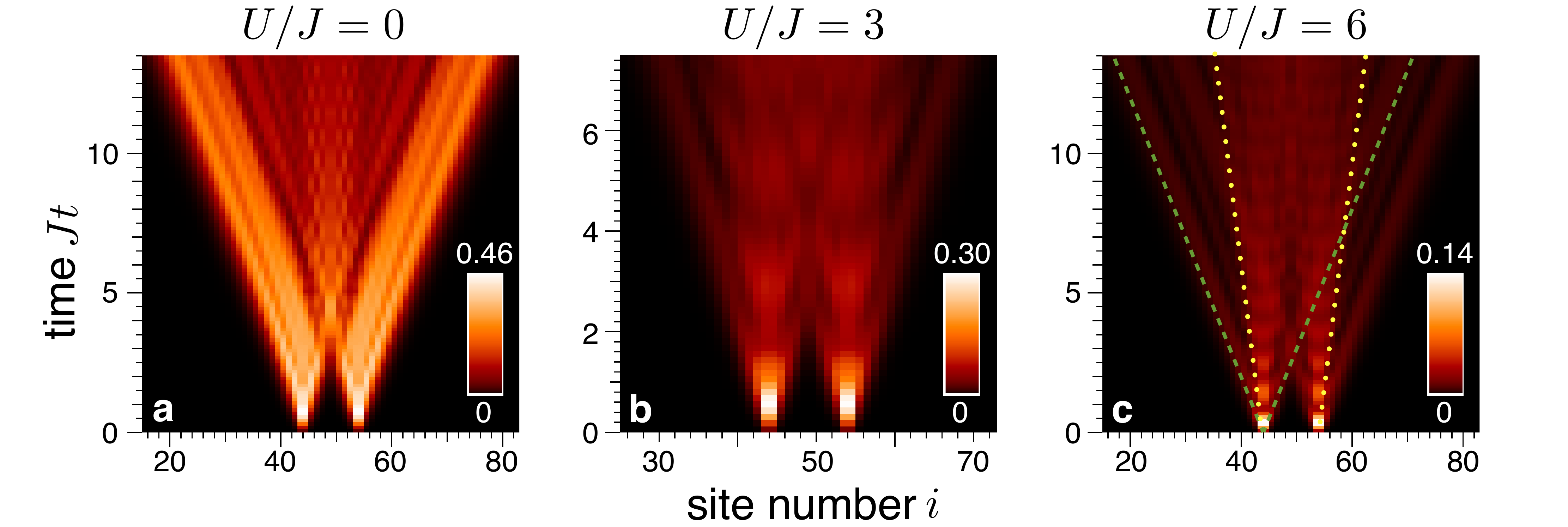}
\par\end{raggedleft}

\caption{Spin-entanglement of a single lattice site $i$ with all the other
sites. The panels show the time evolution of $C_{tot,i}(t)$, given
by equation (\ref{eq:summed concurrence one site}), for different
interaction strengths $U/J$. (a) In the absence of interaction, it
is primarily the sites at the edge of the cloud which are spin-entangled.
(b,c) For finite interactions, we observe the following: At larger
times $C_{tot,i}(t)$ is nearly homogeneous in a central region within
the doublon light cones {[}dotted lines in (c){]}. Spin-entanglement
for lattice sites removed from this central region is finite at locations
corresponding to trajectories of fermions, which have dissolved from
the edges of the cluster. For the left edge the single fermion light
cone is shown as dashed lines. \label{fig:Summed Concurrence one site}}

\end{figure}

Let us consider the spin-entanglement of a site with all the other
lattice sites. The summed concurrence for site $i$ is defined by

\begin{equation}
C_{tot,i}(t)=\sum_{j\neq i}\langle\hat{n}_{i}^{s}(t)\hat{n}_{j}^{s}(t)\rangle\, C_{i,j}(t).\label{eq:summed concurrence one site}\end{equation}
The time evolution of $C_{tot,i}(t)$ is shown in figure~\ref{fig:Summed Concurrence one site}
for different interaction strengths $U/J$. For noninteracting fermions
{[}figure~\ref{fig:Summed Concurrence one site}(a){]}, lattice sites
close to the edge of the cloud display the strongest entanglement,
while sites in the rest of the cloud are hardly entangled. For increasing
interaction strengths a central region with almost uniform $C_{tot,i}(t)$
builds up during the evolution, see figures~\ref{fig:Summed Concurrence one site}(b)
and (c). For large $U/J$, we find that $C_{tot,i}(t)$ approaches
the expectation value $\langle\hat{n}_{i}^{s}\rangle$. This turns
out to be related to the fact that in this case the probability of
having two doublons decay is negligible, and the contribution comes
almost entirely from the decay of a single doublon.

\begin{figure}
\begin{raggedleft}
\includegraphics[scale=0.28]{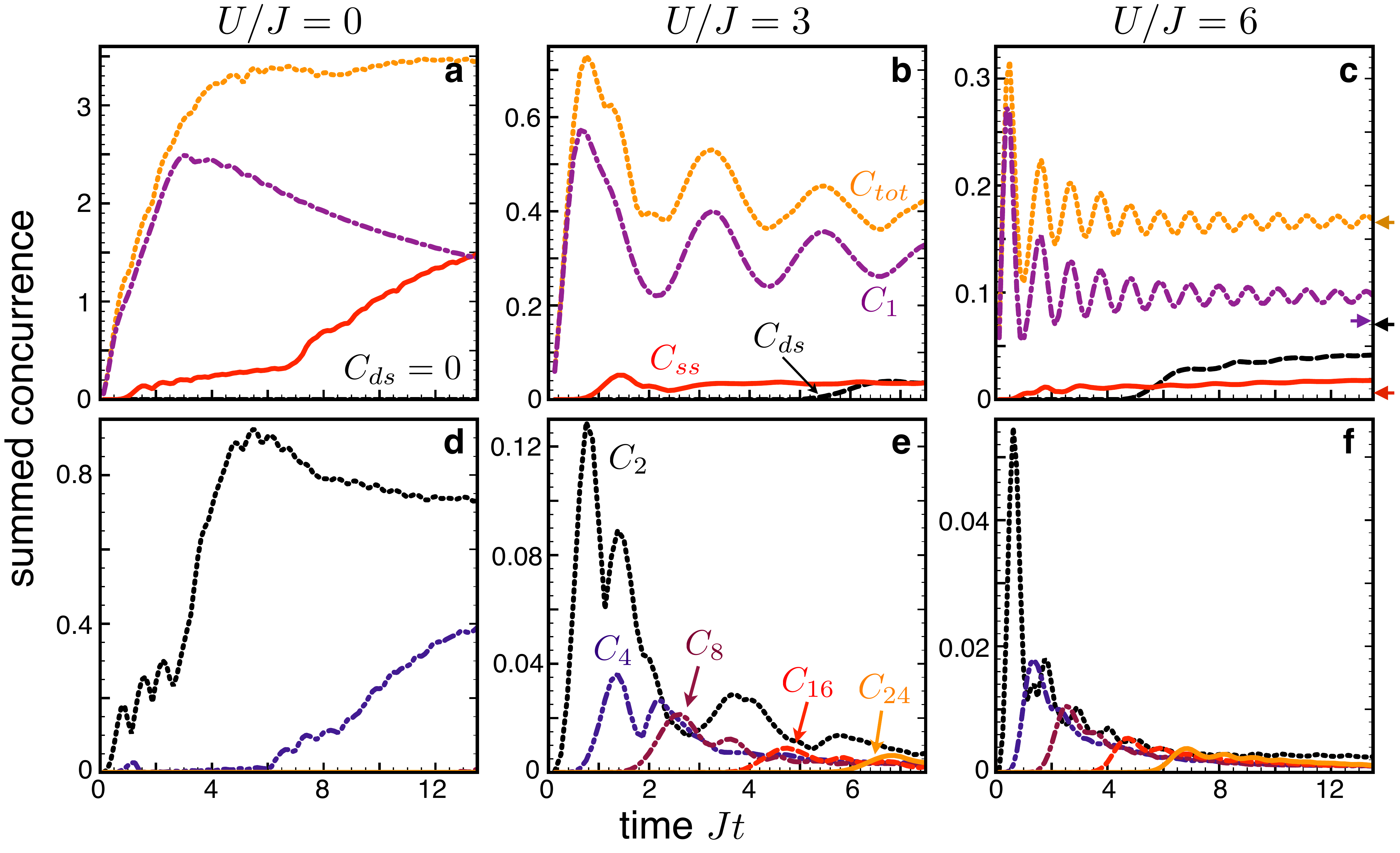}
\par\end{raggedleft}

\caption{Time evolution of spin-entanglement in different areas of the expanding
cloud. (a-c) We show: the summed concurrence $C_{tot}$ of all lattice
sites, the summed concurrence $C_{1}$ between nearest-neighbor sites,
as well as the summed concurrences of all lattice sites at same side
($C_{ss}$) and at different sides ($C_{ds}$) of the cloud (see the
main text for the definitions). Nonvanishing $C_{ds}$ is found only
for finite onsite interaction. $C_{tot}$ and $C_{1}$ quickly settle
into a damped oscillation around a constant value, for $U/J\gtrsim2$.
The arrows in panel (c) show corresponding values of the summed concurrences
for the decay of a single doublon, cf. figure~\ref{fig:Comparison summed concurrence different size of cluster}.
(d-f) Summed concurrence $C_{d}$ of lattice sites at distance $d\geq2$.
(d) Without interactions only close lattice sites are spin-entangled.
(e,f) With increasing interaction strength $U/J$, $C_{d}(t)$ equals
zero for times up to $Jt\approx d/4$, followed by a peak and a decay
for larger times. At fixed time $C_{d}$ is approximately uniform
for the distances $d\lesssim4Jt$.\label{fig:Summed Concurrence different locations}}

\end{figure}

In figure~\ref{fig:Concurrence-unpaired-fermions}(i) it is apparent
that onsite interactions can lead to spin entanglement across the
expanding cluster. In the following we compare the summed concurrences
of lattice sites on the same side and on different sides of the initial
cluster location, $C_{ss}$ and $C_{ds}$, respectively. They are
defined by

\begin{eqnarray}
C_{ss}(t) & = & \sum_{i+1<j<l\:\vee\: i-1>j>r}\langle\hat{n}_{i}^{s}(t)\hat{n}_{j}^{s}(t)\rangle\, C_{i,j}(t)\label{eq:summed concurrence same side}\\
C_{ds}(t) & = & \sum_{i<l\:\wedge\: j>r}\langle\hat{n}_{i}^{s}(t)\hat{n}_{j}^{s}(t)\rangle\, C_{i,j}(t),\label{eq:summed concurrence different side}\end{eqnarray}
where $l$ and $r$ denote the leftmost and rightmost occupied lattice
sites of the initial state. Note that nearest-neighbor lattice sites
are excluded from $C_{ss}(t)$. That means we do not take into account
contributions from virtual transitions of doublons (decaying virtually
into two adjacent fermions) that move away from the cluster initial
position. In addition, we consider the summed concurrence of sites
at fixed distance $d$, $C_{d}(t)=\sum_{i}\langle\hat{n}_{i}^{s}(t)\hat{n}_{i+d}^{s}(t)\rangle C_{i,i+d}(t)$,
and the summed concurrence of all sites, $C_{tot}(t)=\sum_{i<j}\langle\hat{n}_{i}^{s}(t)\hat{n}_{j}^{s}(t)\rangle C_{i,j}(t)$.
The time evolution of these summed concurrences is shown in figure~\ref{fig:Summed Concurrence different locations}. 

For \emph{noninteracting} fermions and small evolution times, the
total summed concurrence is dominated by contributions from nearest-neighbor
sites. For larger times more and more spin-entanglement is transferred
to fermions found on the same side of the initial cluster position
($C_{ss}$), see figure~\ref{fig:Summed Concurrence different locations}(a).
The spin-entanglement remains relevant only for small distances, reflected
in $C_{8}(t)=0$ for all simulated times, cf. figure~\ref{fig:Summed Concurrence different locations}(d). 

By contrast, in the \emph{interacting} case, spin-entanglement is
generated via fermions propagating away on different sides of the
cluster. This is seen as a finite value of $C_{ds}(t)$ for times
$Jt\gtrsim5$, which is the time a hole needs to propagate through
the cluster, cf. figures~\ref{fig:Summed Concurrence different locations}(b)
and (c). The total summed concurrence and concurrence between nearest-neighbor
sites quickly settle into a damped oscillation around a constant value.
The time evolution of the summed concurrence of sites at distance
$d\geq2$ is displayed in figures~\ref{fig:Summed Concurrence different locations}(e)
and (f). $C_{d}(t)$ remains zero for small times, peaks at times
$Jt$ slightly exceeding $d/4,$ and decreases for larger times. This
shows that spin-entangled pairs mainly propagate at almost maximal
(relative) velocity $4J$. For large $U/J$, $C_{d}(t)$ is approximately
uniform for distances $d\leq4Jt$ and roughly decays as $(Jt)^{-1}$
for the simulated times. Note that $C_{1}(t)$ plays a special role.
Its main contribution does not stem from {}``free'' fermions. Rather,
it is generated by a doublon virtually dissolving into adjacent fermions.

\begin{figure}
\begin{centering}
\includegraphics[scale=0.28]{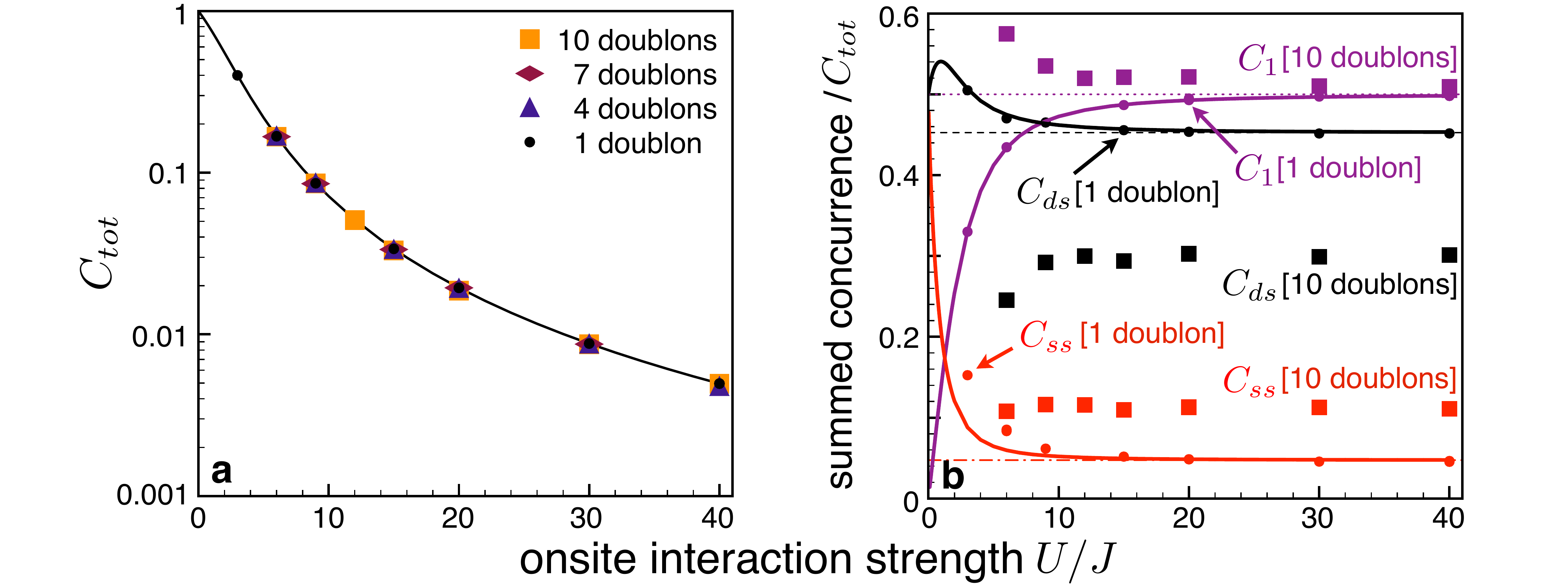}
\par\end{centering}

\caption{Summed concurrences for different sizes of the initial cluster. We
compute the values at time $Jt=20$ (except for $U/J=6$, where $Jt=13.5$),
when they are almost constant as function of time. (a) The total summed
concurrence, $C_{tot},$ shows no dependence on the cluster size for
large onsite interaction strengths $U/J\geq6$. The data agrees with
the exact result for a single doublon derived in appendix~C, $C_{tot}=1-[1+16J^{2}/U^{2}]^{-1/2}$,
which is shown as solid line. (b) The summed concurrences of lattice
sites at same side ($C_{ss}$) and at different sides ($C_{ds}$)
of the cloud (see the main text for the definitions) disagrees for
a single doublon (dots) and a cluster of doublons (squares). Note
that clusters of sizes 4,7, and 10 give similar values. For strong
interactions, a cluster prefers the emission of (delocalized) singlet
pairs into the same direction compared to a single doublon. The summed
concurrence between nearest-neighbor sites ($C_{1}$) approaches $C_{tot}/2$
in both cases. Solid lines show analytical results for $C_{1}$ as
well as the contribution of scattering states to $C_{ss}$ and $C_{ds}$
for a single doublon (see appendix C). \label{fig:Comparison summed concurrence different size of cluster}}

\end{figure}
Let us finally discuss the impact of the cluster size on the spin-entanglement
dynamics. For very weak interactions, a larger number of doublons
means that more delocalized singlet pairs are created shortly after
switching off the confining potential. For large interaction strengths
up to $U/J=40$, we simulate the expansion and compare the summed
concurrences for different cluster sizes, including the case of a
single doublon. We summarize the results in figure~\ref{fig:Comparison summed concurrence different size of cluster}.
Note that reasonably large evolution times ($Jt\approx20$) for the
comparison of the summed concurrences are reached only for $U/J\gtrsim6$.
The summed concurrence of all sites, $C_{tot}$, agrees for all considered
cluster sizes and matches the analytical result for a single doublon,
see figure~\ref{fig:Comparison summed concurrence different size of cluster}(a).
Apparently, the initially Pauli-blocked core has no effect on the
number of created single fermions for the considered times. For $C_{ss}$
and $C_{ds}$, however, we find a clearly different behaviour for
a single doublon and a cluster of four and more doublons {[}figure~\ref{fig:Comparison summed concurrence different size of cluster}(b){]}:
A cluster is more likely to emit (delocalized) spin-entangled pairs
into the same direction.

\subsection{Expansion with modulated tunneling amplitude}

\begin{figure}
\begin{raggedleft}
\includegraphics[scale=0.28]{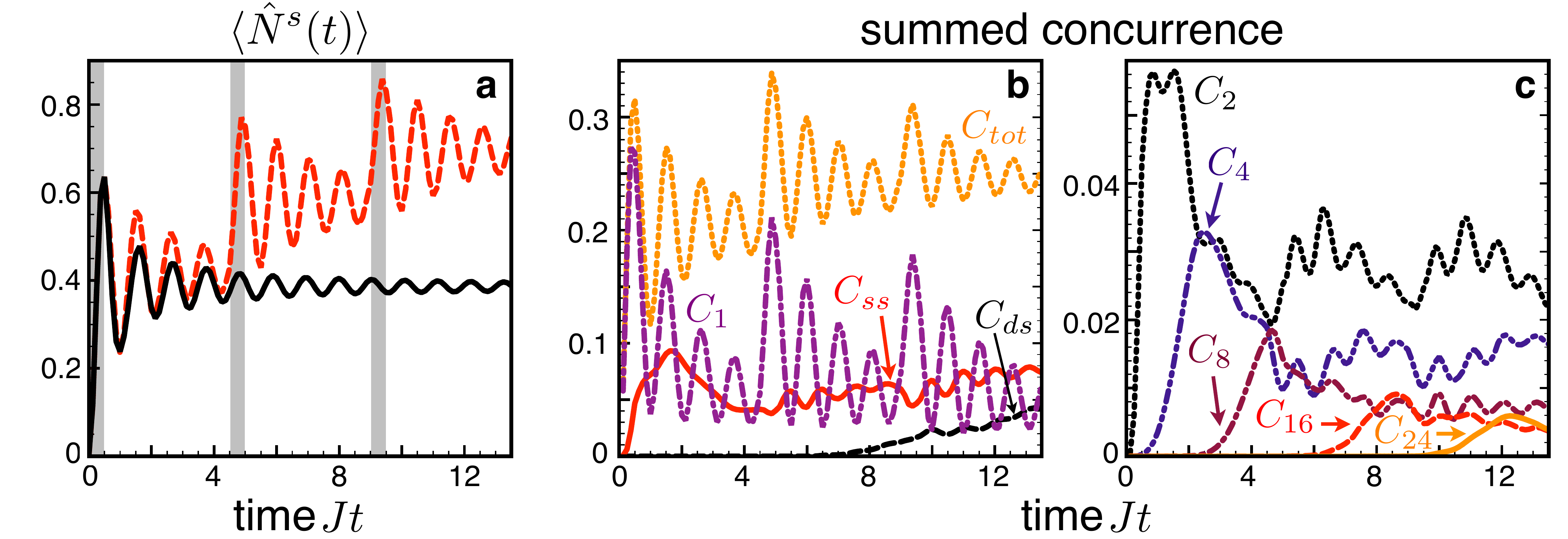}
\par\end{raggedleft}

\caption{Expansion with time-dependent tunneling amplitude (modulated in a
stepwise fashion). The tunneling amplitude is repeatedly switched
between the values $J$ {[}time intervals marked by gray background
in panel (a){]} and $J'=J/2$ {[}white background in panel (a){]},
while the onsite interaction strength is fixed at $U/J=6$. (a) This
dynamics (dashed line) produces a larger total number of single fermions
$\langle\hat{N}^{s}(t)\rangle$ than the free expansion with amplitude
$J$ (solid line). (b,c) Time evolution of the summed concurrences.
In comparison to the free expansion, cf. figures~ \ref{fig:Summed Concurrence different locations}(c)
and (f), the total summed concurrence as well as the concurrences
$C_{ss}$ and $C_{ds}$ are enhanced, while the summed concurrence
at nearest-neighbor sites, $C_{1}$, is decreased. This is also seen
in the enlarged summed concurrence of sites at distances $d\geq2$.
\label{fig:Concurrence-with-modulated-tunneling-maplitude}}

\end{figure}

In the previous section we have seen for the interacting case that
total summed concurrence, $C_{tot},$ approaches a fixed value shortly
after switching off the potential, via the escape of a few fermions
from the cluster edges. Here, we discuss a way of {}``continuously
'' generating single fermions and enhancing $C_{tot}$ compared to
the free time evolution. We consider an expansion during which the
tunneling amplitude is repeatedly varied in time, while the interaction
strength is constant. Such modulation may be experimentally realized
by either varying the laser intensity (the tunneling amplitude decreases
much faster with increased laser intensity than the onsite interaction
strength, see, e.g., \citep{Bloch2008}) or by shaking the lattice
sinusoidally \citep{Lignier2007}. We find for certain values of the
tunneling amplitudes and time intervals between the quenches an increased
amount of spin-entanglement as shown in figure~\ref{fig:Concurrence-with-modulated-tunneling-maplitude}.
For the presented case, the repeated quenches lead to the generation
of more and more single fermions, see figure~\ref{fig:Concurrence-with-modulated-tunneling-maplitude}(a).
This results in an enhanced spin-entanglement between distant lattice
sites, while the spin-entanglement between nearest-neighbor sites
is suppressed {[}compare figures~\ref{fig:Concurrence-with-modulated-tunneling-maplitude}(b)
and \ref{fig:Concurrence-with-modulated-tunneling-maplitude}(c) with
figures~\ref{fig:Summed Concurrence different locations}(c) and
\ref{fig:Summed Concurrence different locations}(f){]}. 

\section{Remarks on observing the spin-entanglement in experiments\label{sec:Experimental remarks}}

In the main part of this article we have analyzed the dynamics of
the spin-entanglement for the expansion from a cluster of doublons.
As discussed in section~\ref{sub: reduced density matrix and concurrence}
the concurrence between two lattice sites $i$ and $j$ can be determined
by the single fermion expectation value $\langle\hat{n}_{i}^{s}(t)\hat{n}_{j}^{s}(t)\rangle$
and the spin-spin correlation $\langle\hat{S}_{i}^{z}(t)\hat{S}_{j}^{z}(t)\rangle=\frac{1}{4}\langle[\hat{n}_{i,\uparrow}(t)-\hat{n}_{i,\downarrow}(t)][\hat{n}_{j,\uparrow}(t)-\hat{n}_{j,\downarrow}(t)]\rangle$. 

Experimentally, both expectation values could be obtained by averaging
over many snapshots of the spin-dependent single-site fermionic particle
number analogous to the already implemented single-site detection
of bosonic particles \citep{Bakr2009,Sherson2010}. Since doubly occupied
lattice sites do not contribute to these expectation values, it would
suffice to be able to detect singly occupied sites. Thus, the loss
of atom pairs due to inelastic light-induced collision during the
imaging process, showing up in the bosonic case, would not be an issue.

Measuring correlations of the spin-$z$ component will be sufficient
for obtaining the concurrence under the following assumptions: (i)
the initial state is of the type we have described (total singlet,
i.e. total spin 0); (ii) the dynamics proceeds according to the model
Hamiltonian, i.e., a SU(2) symmetric Hamiltonian with no decoherence
or entanglement with external degrees of freedom. A scenario where
the initial state violates condition (i) would be a cluster with impurity
sites that contain only single fermions. If there is only one impurity
containing a single fermion, then the problem can still be diagnosed
because the final snapshot will reveal unequal numbers of spin-up
and spin-down fermions. In general, however, if these two conditions
are in doubt, the experiment would have to measure the full spin-spin
density matrix of two lattice sites, by repeated runs and measurements
of spin projections in different directions. This could be done by
implementing a rotation in spin space before measurements, in analogy
to the state tomography realized with trapped ions \citep{Roos2004}.
While being more challenging than measuring the spin-dependent fermion
number, such a kind of coherent spin control seems to be possible
in future experiments. Spin flips at single lattice sites have already
been shown experimentally for ultracold bosons \citep{Weitenberg2011}.
This setup could be in principle extended to coherent spin control,
replacing the Rabi frequency sweep by driving a Rabi oscillation \citep{Weitenberg2011a}.

\section{Summary and Outlook\label{sec:Summary-and-Outlook}}

In this article, we have analyzed the creation and time-evolution
of spin-entanglement during the sudden expansion from a cluster of
doublons into an empty lattice. Interestingly, remote spin-entangled
pairs are created for large onsite interaction. In addition, an extended
cluster favours the emission of the two fermions of a pair into the
same direction when compared against the decay of a single doublon.
Finally, we found that a time-dependent modulation of the tunneling
amplitude can be used to increase the {}``production'' of spin-entangled
pairs. Our results provide a starting point for studying the spin-entanglement
dynamics for more complex initial states, in different dimensionalities,
e.g., the crossover from one to two dimensions, or for spin-imbalanced
fermionic gases.

In the scenario considered here, the spin-entanglement can be extracted
from the two-site spin-$z$ correlation functions. Thus, it will become
experimentally accessible once spin-dependent single-site detection
has been implemented.

\section*{Acknowledgment}

\addcontentsline{toc}{section}{Acknowledgment}

We are grateful to Fabian Heidrich-Meisner for valuable feedback.
We thank the DFG for support in the Emmy-Noether programme and the
SFB/TR 12. Ian McCulloch acknowledges support from the Australian
Research Council Centre of Excellence for Engineered Quantum Systems
and the Discovery Projects funding scheme (Project No. DP1092513).

\appendix

\section*{Appendix A. Correlation functions and concurrence of noninteracting
fermions}

\addcontentsline{toc}{section}{Appendix A. Correlation functions and concurrence of noninteracting fermions}
\setcounter{section}{1}

In this appendix, we derive analytical formulae for the correlation
functions of expanding, \emph{noninteracting} fermions. For vanishing
onsite interaction, fermions of different species are uncorrelated
and each fermion propagates with the free dispersion. The time evolution
of the annihilation operators is given by \begin{equation}
\hat{c}_{j,a}(t)=\sum_{m}G_{jm}(t)\hat{c}_{m,a},\end{equation}
with spin index $a=\uparrow,\downarrow$ and the free fermion propagator

\begin{equation}
G_{jm}(t)=\int_{-\pi}^{\pi}\frac{\mbox{d}k}{2\pi}\exp\{i[(m-j)k-2J\cos(k)t]\}=i^{j-m}\mathcal{J}_{j-m}(2Jt).\end{equation}
Here, $\mathcal{J}$ denotes the Bessel function of the first kind
and we assumed an infinite lattice. For the previously discussed initial
state of localized doublons, the Green's function reads 

\begin{equation}
\mathcal{G}_{ij}(t)=\langle\hat{c}_{i,a}^{\dagger}(t)\hat{c}_{j,a}(t)\rangle=i^{j-i}\sum_{m\in O}\mathcal{J}_{j-m}(2Jt)\mathcal{J}_{i-m}(2Jt),\end{equation}
where the summation is taken over all initially occupied lattice sites
$O$. For both spin species, the density is given by 

\begin{equation}
\mathcal{N}_{i}(t)=\mathcal{G}_{ii}(t)=\sum_{m\in O}\mathcal{J}_{|i-m|}^{2}(2Jt).\end{equation}

When the fermions are initially localized at the lattice sites, equal
time, normal ordered products of operators can be expressed as a Slater
determinant of the equal time one-particle Green's functions, $\mathcal{G}_{ij}(t)$,
\citep{Loewdin1955}. The correlation function $\langle\hat{S}_{i}^{z}(t)\hat{S}_{j}^{z}(t)\rangle$
can be evaluated by writing it in terms of density-density correlations
and making use of $\langle\hat{n}_{i,a}(t)\hat{n}_{j\neq i,b}(t)\rangle=\mathcal{N}_{i}(t)\mathcal{N}_{j}(t)-\delta_{ab}\left|\mathcal{G}_{ij}(t)\right|^{2}$
and $\langle\hat{n}_{i,a}(t)\hat{n}_{i,b}(t)\rangle=\mathcal{N}_{i}^{(2-\delta_{ab})}$.
This yields\begin{eqnarray}
\langle\hat{S}_{i}^{z}(t)\hat{S}_{j\neq i}^{z}(t)\rangle & = & -\frac{1}{2}\left|\mathcal{G}_{ij}(t)\right|^{2},\label{eq:spin-spin correlation noninteracting case}\\
\langle\hat{S}_{i}^{z}(t)\hat{S}_{i}^{z}(t)\rangle & = & \frac{1}{2}\mathcal{N}_{i}(t)\left[1-\mathcal{N}_{i}(t)\right].\end{eqnarray}
Note that the probability of finding two fermions with the same spin
at lattice sites $i$ and $j$ is by $2|\mathcal{G}_{ij}(t)|^{2}$
smaller than the probability for fermions with antiparallel spin.
This difference leads to a nonvanishing spin-spin correlation even
in the absence of onsite interactions. For noninteracting fermions,
the spin-spin correlation is related to the density-density correlation
$D_{ij}(t)=\langle\hat{n}_{i}(t)\hat{n}_{j}(t)\rangle-\langle\hat{n}_{i}(t)\rangle\langle\hat{n}_{j}(t)\rangle$
via

\begin{equation}
D_{ij}(t)=4\langle\hat{S}_{i}^{z}(t)\hat{S}_{j}^{z}(t)\rangle.\label{eq:relation of spin and density correlation noninteracting case}\end{equation}
Thus, the density-density correlation $D_{ij}(t)$ is smaller or equal
to zero for different lattice sites $i$ and $j$, cf. figures~\ref{fig:density-density-fluctuation}(a)
and (c)-(e). Analogously, the density-density correlation of single
fermions is calculated using the relations for $\langle\hat{n}_{i,a}(t)\hat{n}_{j,b}(t)\rangle$
given above. We find\begin{eqnarray}
\langle\hat{n}_{i}^{s}(t)\hat{n}_{i}^{s}(t)\rangle & = & 2\mathcal{N}_{i}(t)\left[1-\mathcal{N}_{i}(t)\right],\\
\langle\hat{n}_{i}^{s}(t)\hat{n}_{j\neq i}^{s}(t)\rangle & = & 4\left[N_{i}(t)-\left(\mathcal{N}_{i}(t)\mathcal{N}_{j}(t)-\left|\mathcal{G}_{ij}(t)\right|^{2}\right)\right]\times\label{eq:single fermion density correlation noninteracting case}\\
 &  & \left[N_{j}(t)-\left(\mathcal{N}_{i}(t)\mathcal{N}_{j}(t)-\left|\mathcal{G}_{ij}(t)\right|^{2}\right)\right]-2\left|\mathcal{G}_{ij}(t)\right|^{2}.\nonumber \end{eqnarray}
The concurrence $C_{i,j}(t)$ {[}defined by equation~(\ref{eq:Concurrence}){]}
can be evaluated using the expressions~(\ref{eq:spin-spin correlation noninteracting case})
and (\ref{eq:single fermion density correlation noninteracting case}).
It approaches one in the limit $\langle\hat{S}_{i}^{z}(t)\hat{S}_{j\neq i}^{z}(t)\rangle/\langle\hat{n}_{i}^{s}(t)\hat{n}_{j\neq i}^{s}(t)\rangle\searrow-\frac{1}{4}$.
This is the case for $\left|\mathcal{G}_{ij\neq i}(t)\right|^{2}\nearrow\mathcal{N}_{i}(t)\mathcal{N}_{j\neq i}(t)$
(note that $\left|\mathcal{G}_{ij\neq i}(t)\right|^{2}$ can only
assume values between $0$ and $1/2$).

\appendix

\section*{Appendix B. Decay of a doublon into scattering states}

\addcontentsline{toc}{section}{Appendix B. Decay of a doublon into scattering states}
\setcounter{section}{2}

This appendix presents the derivation of the decay probability of
a fermionic doublon into different scattering states $|\psi_{K,k}\rangle$.
In doing so, the wavefunctions of the scattering states are calculated
mainly adopting the procedure for two-particle states in the Bose
Hubbard model presented in \citep{Valiente2008}.

For one spin-up and one spin-down fermion, and an infinite lattice
the Hamiltonian~(\ref{eq:FH Hamiltonian}) can be expressed in the
form \begin{eqnarray}
\hat{\mathcal{H}}_{D} & = & -J\sum_{i}\left\{ \ket{\uparrow_{i}}\bra{\uparrow_{i+1}}+\ket{\uparrow_{i+1}}\bra{\uparrow_{i}}+\ket{\downarrow_{i}}\bra{\downarrow_{i+1}}+\ket{\downarrow_{i+1}}\bra{\downarrow_{i}}\right\} \nonumber \\
 &  & +U\sum_{i}\ket{\uparrow_{i},\downarrow_{i}}\bra{\uparrow_{i},\downarrow_{i}}.\label{eq:Hubbard Hamiltonian two-particle sector}\end{eqnarray}
Analogously, we write the two-fermion states in terms of the basis
$\{\ket{\uparrow_{i},\downarrow_{j}}\}$, $|\Psi\rangle=\sum_{i,j}\Psi(\uparrow_{i},\downarrow_{j})\ket{\uparrow_{i},\downarrow_{j}}.$
For the decay of a doublon, the fermions are always in the spin-singlet
subspace. Thus, we can write the two-particle wavefunction as $\Psi(\uparrow_{i},\downarrow_{j})=\varphi_{s}(\uparrow,\downarrow)\cdot\psi(i,j)$,
with the spin-singlet wavefunction $\varphi_{s}(a,b)=\delta_{a,\uparrow}\delta_{b,\downarrow}-\delta_{a,\downarrow}\delta_{b,\uparrow}$
and a symmetric spatial wavefunction, $\psi(i,j)=\psi(j,i)$. Plugging
$\ket{\Psi}$ into the Schr\"odinger equation for Hamiltonian (\ref{eq:Hubbard Hamiltonian two-particle sector})
yields\begin{eqnarray}
\left(E-U\delta_{ij}\right)\psi(i,j) & =-J & \left[\psi(i-1,j)+\psi(i+1,j)+\psi(i,j-1)\right.\nonumber \\
 &  & \left.+\psi(i,j+1)\right].\end{eqnarray}

This relation is simplified by introducing center of mass (c.o.m.)
and relative coordinates, $R=(i+j)/2$ and $r=i-j$, respectively.
The wavefunction factorizes into a plane wave motion of the c.o.m.
with total wavenumber $K\in[-\pi,\pi)$ and a $K$-dependent relative
motion, i.e., $\psi(i,j)=e^{iKR}\psi_{K}(r)$. The relative motion
satisfies the recurrence relation 

\begin{equation}
-J_{K}\left[\psi_{K}(r-1)+\psi_{K}(r+1)\right]=\left(E_{K}-U\delta_{r0}\right)\psi_{K}(r),\label{eq:SEQ for relative motion}\end{equation}
with $K$-dependent tunneling amplitude $J_{K}=2J\cos(K/2)$ and energy
$E_{K}$. For vanishing interaction strength $U$ equation~(\ref{eq:SEQ for relative motion})
is solved by plane waves $\psi_{K,k}(r)=e^{\pm ikr}$ with corresponding
eigenenergies $E_{K,k}=-2J_{K}\cos(k)$. In the interacting case,
we make a scattering ansatz by writing the wavefunction as a superposition
of incoming and reflected plane waves, $\psi_{K,k}(r\geq0)=e^{ikr}+c\, e^{-ikr}$.
Here, $k$ is the relative wavenumber and $\psi_{K,k}(r<0)$ is determined
by the symmetry condition $\psi(r)=\psi(-r)$. The boundary condition
at $r=0$ in equation~ (\ref{eq:SEQ for relative motion}) fixes
the coefficient $c$ and we obtain

\begin{equation}
\psi_{K,k}(r)=\psi_{K,k}(0)\ \left[\cos(kr)+\frac{U}{2J_{K}\sin(k)}\sin(k|r|)\right].\end{equation}
Finally, we compare the decay probability of a doublon for different
scattering states $\ket{\psi_{K,k}}$. In doing so, we express $|\psi_{K,k}(0)|^{2}$
in terms of the average density in the relative coordinate, $\bar{n}$,
which is obtained by averaging $|\psi_{K,k}(r)|^{2}$ over one period
$2\pi/k$. Note that $\bar{n}$ does only depend on the systems size
and is independent of $k,K,$ and $U$. We find that the decay probability
into the scattering state $\ket{\psi_{K,k}}$ equals

\begin{equation}
|\psi_{K,k}(0)|^{2}=\bar{n}\,\left[1+U^{2}/\left(16J^{2}\cos^{2}(K/2)\sin^{2}(k)\right)\right]^{-1}.\label{eq:squared overlap of doublon and scattering states}\end{equation}

\appendix

\section*{Appendix C. Summed concurrences of a single doublon}

\addcontentsline{toc}{section}{Appendix C. Summed concurrences of a single doublon}
\setcounter{section}{3}

This appendix provides the calculation of the summed concurrences
for the expansion from a single doublon. The results are depicted
in figure~\ref{fig:Comparison summed concurrence different size of cluster}.
For this initial state, the fermions form a spin-singlet for all times.
Consequently, the concurrence $C_{i,j}(t)$ {[}defined by equation
(\ref{eq:Concurrence}){]} equals one for all sites $i$ and $j$,
and the summed concurrences simplify to sums over the expectation
values $\langle\hat{n}_{i}^{s}(t)\hat{n}_{j}^{s}(t)\rangle$, see
section~\ref{sub:Summed-concurrences}.

Let us first consider the total summed concurrence for a single doublon
$C_{tot}(t)=\sum_{i<j}\langle\hat{n}_{i}^{s}(t)\hat{n}_{j}^{s}(t)\rangle$.
This is nothing but the probability of finding the fermions at different
lattice sites. It can be expressed as $1-P_{D}(t)$, with the doublon
survival probability $P_{D}(t)$, i.e., the probability of finding
the doublon intact at time $t$. In the long time limit, only fermions
in a bound state remain localized close to each other. For a singlet
pair with finite onsite interaction $U\neq0$ and c.o.m. wavenumber
$K$, it exists one bound state $\psi_{K}^{b}(r)$ {[}localized solution
of equation~(\ref{eq:SEQ for relative motion}), details are given
below{]}, where $r$ is the relative coordinate. Thus, we obtain $P_{D}(\infty)=\frac{1}{2\pi}\int_{-\pi}^{\pi}\mbox{d}K\,|\psi_{K}^{b}(0)|^{4}$
in the limit of an infinite lattice, where $\psi_{K}^{b}(0)$ is the
overlap between the doublon and the bound state. Analogously, we find
$C_{1}(\infty)=\frac{1}{2\pi}\int_{-\pi}^{\pi}\mbox{d}K\,|\psi_{K}^{b}(0)|^{2}|\psi_{K}^{b}(1)|^{2}$,
where we have used that $C_{1}(t)=\sum_{i}\langle\hat{n}_{i}^{s}(t)\hat{n}_{i+1}^{s}(t)\rangle$
equals the probability of finding the fermions at time $t$ at nearest-neighbor
lattice sites. The bound state can be calculated using the exponential
ansatz $\psi_{K}^{b}(r)=1/\sqrt{\mathcal{N}_{K}}\cdot\alpha_{K}^{|r|}$
for the wavefunction in equation~(\ref{eq:SEQ for relative motion}).
This gives $\alpha_{K}=U/2J_{K}-\mbox{sign}(U/2J_{K})\cdot\sqrt{1+(U/2J_{K})^{2}}$
and $\mathcal{N}_{K}=(2J_{K}/U)\cdot\sqrt{1+(U/2J_{K})^{2}}$, with
$J_{K}=2\cos(K/2)$. Inserting this solution into the doublon survival
probability yields $P_{D}(\infty)=\frac{1}{2\pi}\int_{-\pi}^{\pi}\mbox{d}K\,\mathcal{N}_{K}^{-2}=[1+16J^{2}/U^{2}]^{-1/2}$.
In the limit of infinite times, the total summed concurrence and summed
concurrence of nearest-neighbor lattice sites are, hence, given by

\begin{eqnarray}
C_{tot}(\infty) & = & 1-[1+16J^{2}/U^{2}]^{-1/2}=8J^{2}/U^{2}+\mathcal{O}([J/U]^{4}),\\
C_{1}(\infty) & = & \frac{1}{2\pi}\int_{-\pi}^{\pi}\mbox{d}K\,|\alpha_{K}|^{2}\mathcal{N}_{K}^{-2}=4J^{2}/U^{2}+\mathcal{O}([J/U]^{4}).\end{eqnarray}

In the long-time limit, the summed concurrences of lattice sites at
same side ($C_{ss}(t)$) and at different sides ($C_{ds}(t)$) of
initial doublon position can be related to the scattering states calculated
in appendix B. We denote by $C_{ds}^{(scat)}$ and $C_{ss}^{(scat)}$
the contributions to the summed concurrences stemming from the scattering
states. $C_{ds}^{(scat)}$ is the sum of the occupation probabilities
of scattering states corresponding to fermions moving in opposite
direction {[}$k_{1}\in(0,\pi)$, $k_{2}\in(-\pi,0)$ or $k_{1}\in(-\pi,0)$,
$k_{2}\in(0,\pi)${]}, and $C_{ss}^{(scat)}$ is the sum of the occupation
probabilities of those states with fermions moving into the same direction
{[}$k_{1},k_{2}\in(0,\pi)$ or $k_{1},k_{2}\in(-\pi,0)${]}. Here,
$k_{1}$ and $k_{2}$ denote the asymptotic wavenumbers of the two
fermions. The occupation probabilities are the absolute square of
the overlap between the doublon and the scattering state, $|\psi_{K,k}(0)|^{2}$.
The explicit form of $|\psi_{K,k}(0)|^{2}$ is given by equation~(\ref{eq:squared overlap of doublon and scattering states}).
Taking the continuum limit we find

\begin{eqnarray}
C_{ds}^{(scat)} & = & \frac{1}{(2\pi)^{2}}\left[\int_{0}^{\pi}\mbox{d}k_{1}\int_{-\pi}^{0}\mbox{d}k_{2}+\int_{-\pi}^{0}\mbox{d}k_{1}\int_{0}^{\pi}\mbox{d}k_{2}\right]\\
 &  & \times\left[1+U^{2}/\left(16J^{2}\cos^{2}([k_{1}+k_{2}]/2)\sin^{2}([k_{1}-k_{2}]/2)\right)\right]^{-1}\nonumber \\
 & = & 4\frac{J^{2}}{U^{2}}[1/2+4/\pi^{2}]+\mathcal{O}([J/U]^{4}),\end{eqnarray}

\begin{eqnarray}
C_{ss}^{(scat)} & = & \frac{1}{(2\pi)^{2}}\left[\int_{0}^{\pi}\mbox{d}k_{1}\int_{0}^{\pi}\mbox{d}k_{2}+\int_{-\pi}^{0}\mbox{d}k_{1}\int_{-\pi}^{0}\mbox{d}k_{2}\right]\\
 &  & \times\left[1+U^{2}/\left(16J^{2}\cos^{2}([k_{1}+k_{2}]/2)\sin^{2}([k_{1}-k_{2}]/2)\right)\right]^{-1}\nonumber \\
 & = & 4\frac{J^{2}}{U^{2}}[1/2-4/\pi^{2}]+\mathcal{O}([J/U]^{4}).\end{eqnarray}
The summed concurrences $C_{ss}(t)$ and $C_{ds}(t)$ used for the
numerical data {[}equations~(\ref{eq:summed concurrence same side})
and (\ref{eq:summed concurrence different side}){]} contain additional
contributions from bound states. In the definition of $C_{ss}(t)$
we excluded nearest-neighbor lattice sites in order to remove most
of these contributions {[}note that nearest-neighbor sites do not
appear in $C_{ds}(t)${]}. From the expression for the bound state
given above follows that all other terms are of the order $\mathcal{O}([J/U]^{4})$.
Thus, $C_{ss}$(t) and $C_{ds}(t)$ agree with $C_{ss}^{(scat)}$
and $C_{ds}^{(scat)}$ for large $Jt$ and $U/J$, cf. figure~\ref{fig:Comparison summed concurrence different size of cluster}(b).

In conclusion, we find following relations between the summed concurrences
for long evolution times and large interaction strengths $U/J\gg1$:
$C_{1}=C_{tot}/2$, $C_{ss}=[1/4-2/\pi^{2}]C_{tot}$, and $C_{ds}=[1/4+2/\pi^{2}]C_{tot}$.

\bibliographystyle{iopart_num}
\addcontentsline{toc}{section}{\refname}\bibliography{Entanglement1,Entanglement2}

\end{document}